\documentclass[aps,prx,amsmath,amssymb,floatfix,twocolumn,amsmath,superscriptaddress,nofootinbib,letterpaper]{revtex4-1}
\usepackage{multirow}
\usepackage{bbold}
\usepackage{subfigure}
\usepackage{mathrsfs}
\usepackage{hyperref}
\usepackage[normalem]{ulem}
\usepackage{bm}

\usepackage[utf8]{inputenc}
\usepackage{amsfonts, relsize, color}
\usepackage{graphics}
\usepackage{graphicx}

\newcommand{\cx}{\cos(k_x a)}
\newcommand{\cy}{\cos(k_y a)}
\newcommand{\cz}{\cos(k_z a)}

\newcommand{\Sx}{\sin(k_x)}
\newcommand{\Sy}{\sin(k_y)}

\newcommand{\Cx}{\cos(k_x)}
\newcommand{\Cy}{\cos(k_y)}
\newcommand{\Cz}{\cos(k_z)}

\begin{document}
\title{\bf Non-Abelian anomalies in multi-Weyl semimetals}

\author{Renato M. A. Dantas}\email{rmad@pks.mpg.de}
\affiliation{Max-Planck-Institut f\"{u}r Physik komplexer Systeme, N\"{o}thnitzer Str. 38, 01187 Dresden, Germany}

\author{Francisco Pe\~{n}a-Benitez}\email{pena@pks.mpg.de}
\affiliation{Max-Planck-Institut f\"{u}r Physik komplexer Systeme, N\"{o}thnitzer Str. 38, 01187 Dresden, Germany}

\author{Bitan Roy}\email{bir218@lehigh.edu}
\affiliation{Max-Planck-Institut f\"{u}r Physik komplexer Systeme, N\"{o}thnitzer Str. 38, 01187 Dresden, Germany}
\affiliation{Department of Physics, Lehigh University, Bethlehem, Pennsylvania 18015, USA}

\author{Piotr Sur\'owka}\email{surowka@pks.mpg.de}
\affiliation{Max-Planck-Institut f\"{u}r Physik komplexer Systeme, N\"{o}thnitzer Str. 38, 01187 Dresden, Germany}

\date{\today}
\begin{abstract}
We construct the effective field theory for time-reversal symmetry breaking multi-Weyl semimetals (mWSMs), composed of a single pair of Weyl nodes of (anti-)monopole charge $n$, with $n=1,2,3$ in crystalline environment. From both the continuum and lattice models, we show that a mWSM with $n>1$ can be constructed by placing $n$ flavors of linearly dispersing simple Weyl fermions (with $n=1$) in a bath of an $SU(2)$ non-Abelian static background gauge field. Such an $SU(2)$ field preserves certain crystalline symmetry (four-fold rotational or $C_4$ in our construction), but breaks the Lorentz symmetry, resulting in nonlinear band spectra (namely, $E \sim (p^2_x + p^2_y)^{n/2}$, but $E \sim |p_z|$, for example, where momenta ${\bf p}$ is measured from the Weyl nodes). Consequently, the effective field theory displays $U(1) \times SU(2)$ non-Abelian anomaly, yielding anomalous Hall effect, its non-Abelian generalization, and various chiral conductivities. The anomalous violation of conservation laws is determined by the monopole charge $n$ and a specific algebraic property of the $SU(2)$ Lie group, which we further substantiate by numerically computing the regular and ``isospin" densities from the lattice models of mWSMs. These predictions are also supported from a strongly coupled (holographic) description of mWSMs. Altogether our findings unify the field theoretic descriptions of mWSMs of arbitrary monopole charge $n$ (featuring $n$ copies of the Fermi arc surface states), predict signatures of non-Abelian anomaly in table-top experiments, and pave the route to explore anomaly structures for multi-fold fermions, transforming under arbitrary half-integer or integer spin representations.     
\end{abstract}

\maketitle

\section{Introduction}

Anomalies are traditionally studied in the realm of relativistic field theories that are pertinent in high-energy physics \cite{Bertlmann:1996xk,Bell:1969ts,Adler,NielsenABJ,fujikawa,KHARZEEV20161,PhysRev.184.1848,BARDEEN1984421}. They show up as the violation of symmetries of the classical action upon quantization of chiral massless fermions. An intrinsic feature of high-energy theories is that they are Lorentz symmetric, stemming from the linear dispersion of the chiral fermions. Also in the world of condensed matter systems an \emph{emergent} relativistic symmetry results from the quasiparticle spectra that are linear in momentum, but at low energies. This is the quintessential feature of Weyl semimetals - a class of materials where quantum anomaly has been studied theoretically~\cite{Jackiw:1999qq,PhysRevB.88.245107,Grushin:2012mt,Rebhan:2009vc,Landsteiner:2013aba,PhysRevB.86.115133,PhysRevX.4.031035,PhysRevB.92.075205} and its signature has possibly been observed in experiments~\cite{Exp1,Exp2,Exp3,Exp4,Exp5,Exp6,Exp7,Exp8,Gooth:2017mbd,Schindler:2018wrd,2017NatCo...813741Z}.

\begin{figure}[t!]
\subfigure[]{
\includegraphics[height=3.80cm]{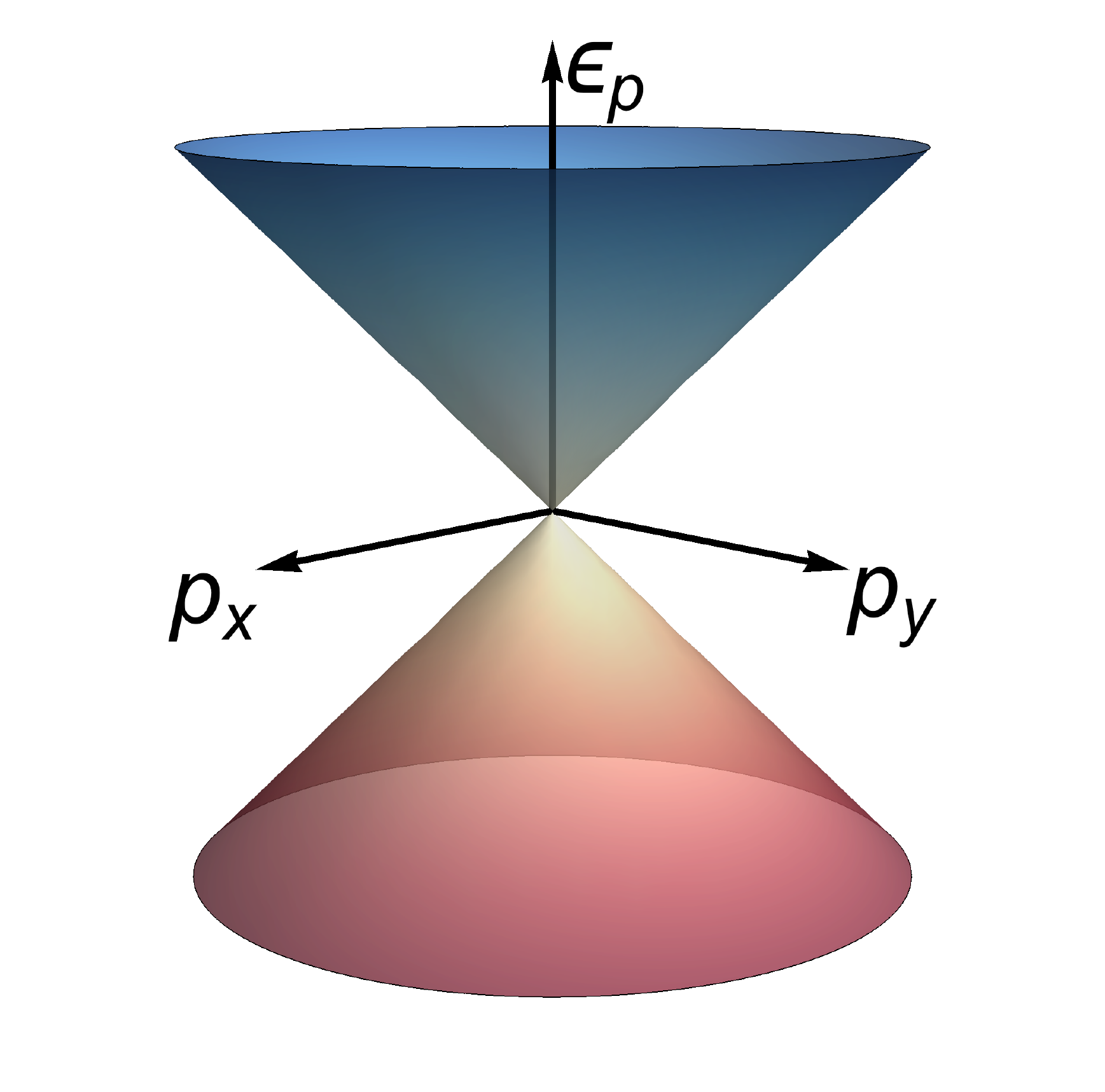}
}%
\subfigure[]{
\includegraphics[height=3.75cm]{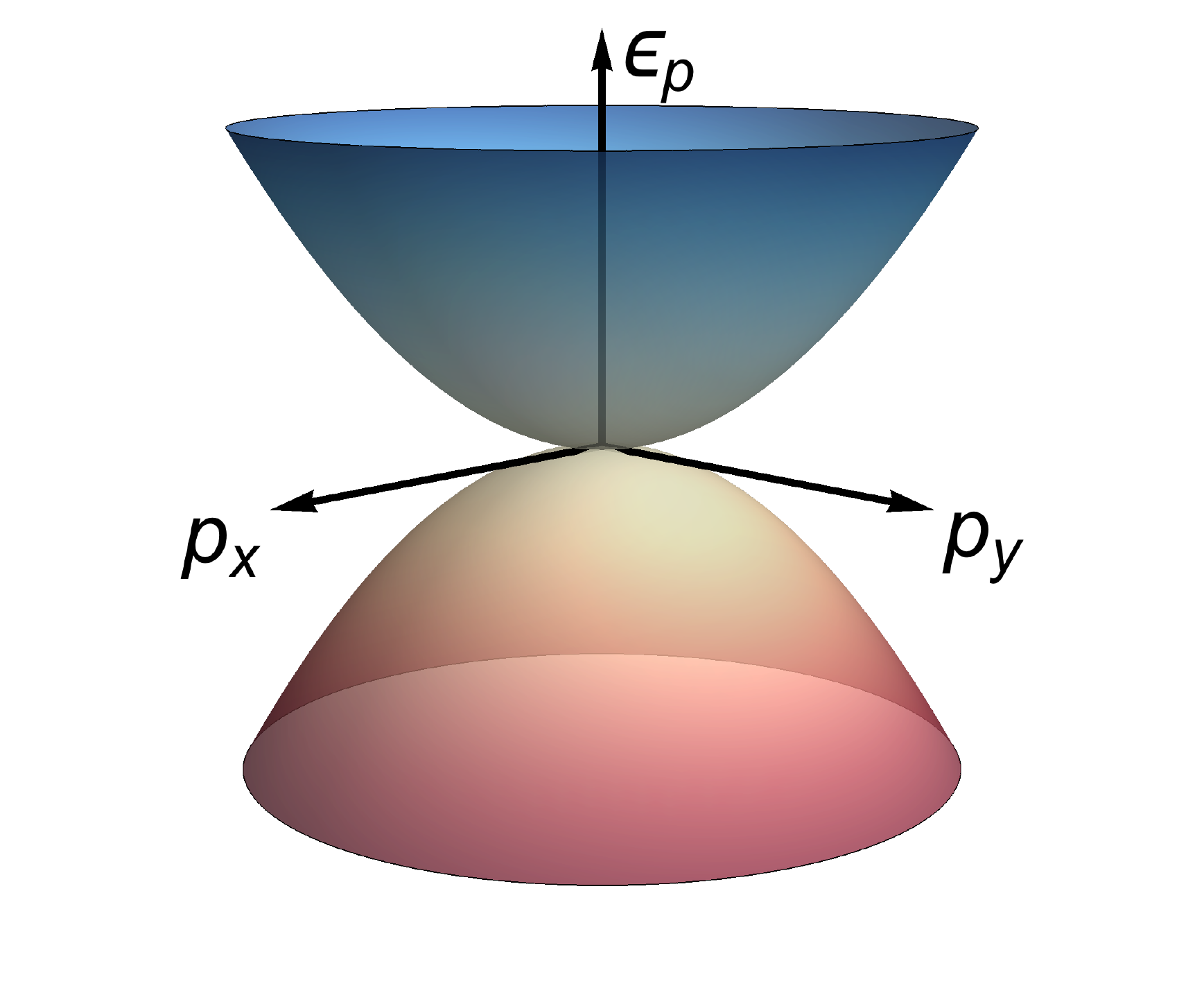}
}
\subfigure[]{
\includegraphics[height=3.60cm]{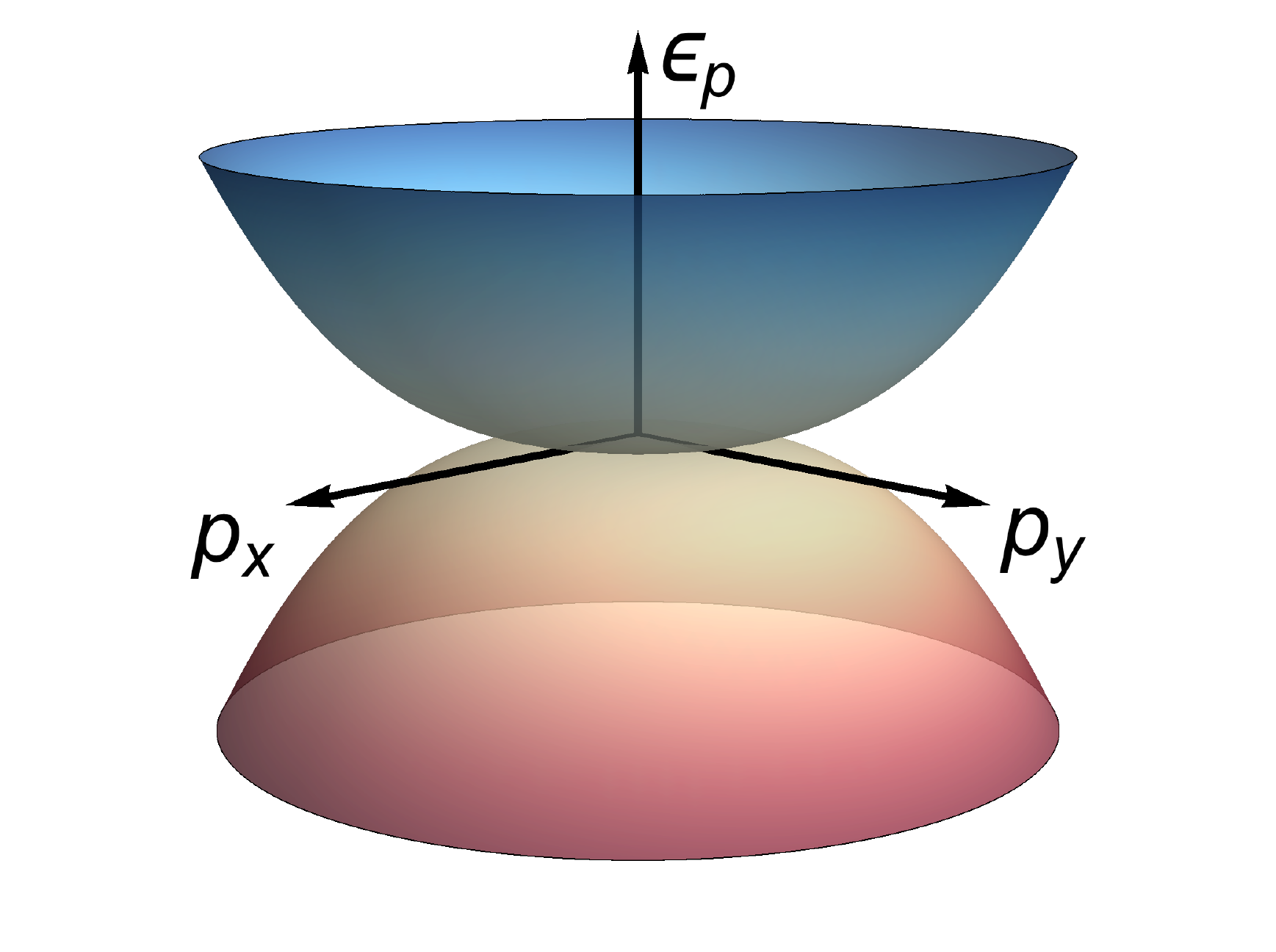}
}%
\subfigure[]{
\includegraphics[height=3.65cm]{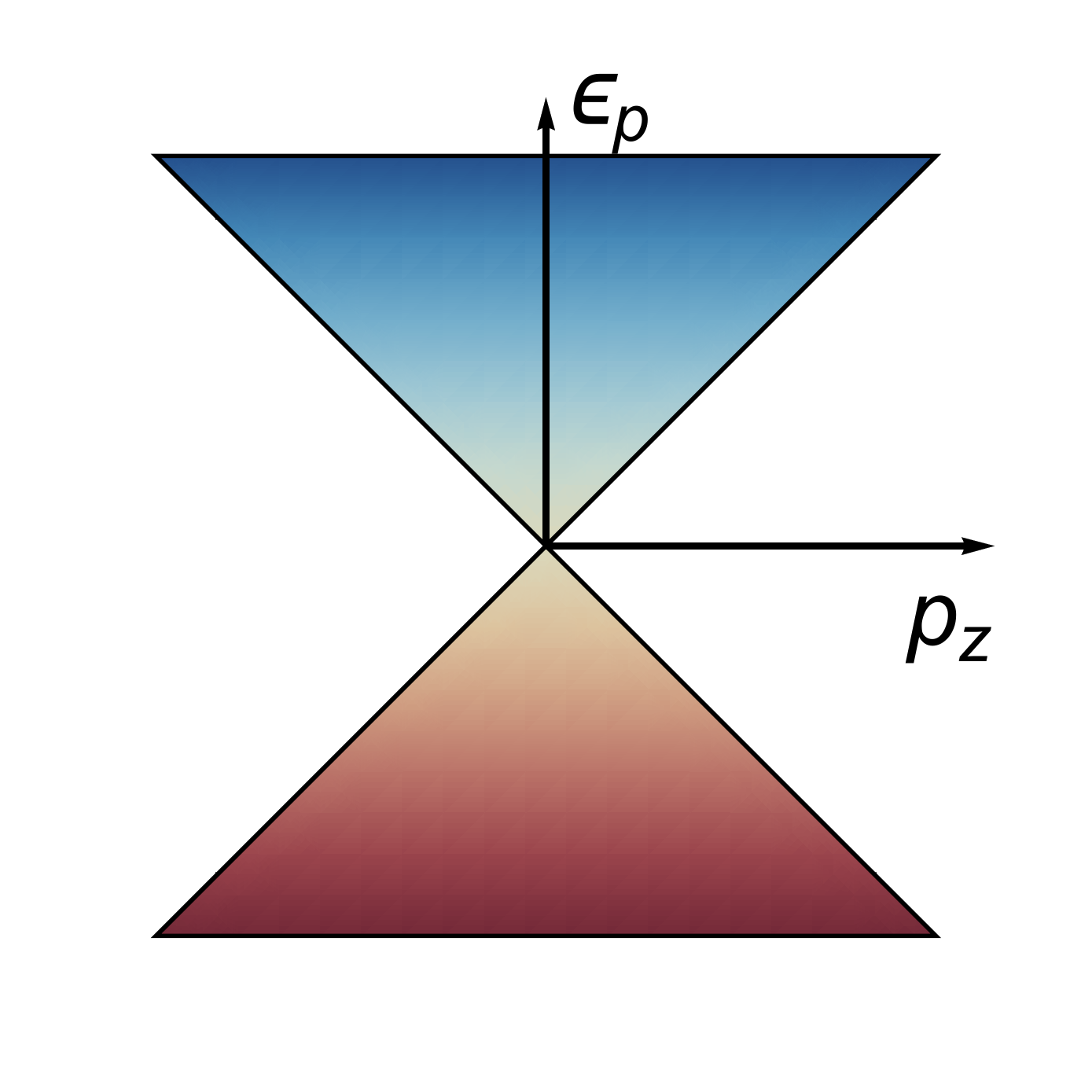}
}%
\caption{ Energy dispersion ($\epsilon_{\bf p}$) for multi-Weyl fermions in the $(p_x, p_y)$ plane for (a) $n=1$, (b) $n=2$ and (c) $n=3$ around a Weyl node. Respectively for $n=1,2$ and $3$ the valence and conduction bands display linear, quadratic and cubic touching in this plane. However, the dispersion always scales linearly with $p_z$ irrespective of $n$, as shown in (d). The momenta ${\bf p}$ is measured from the Weyl nodes, placed at ${\bf p}=0$.
}~\label{Fig:spectra_multiWeyl}
\end{figure}

More intriguingly, condensed matter systems offer unique opportunities to further extend our understanding of anomalies in quantum field theories and its connections with transport. Recent developments have allowed us to go beyond the original paradigm of linearly dispersing chiral fermions, as nowadays gapless chiral systems with finite band curvatures can be found in various solid state compounds. The main motivation of our work is to pedagogically develop a comprehensive understanding of such systems, lacking the Lorentz symmetry from their effective low energy field theory and anchor various field theoretic predictions from concrete, but simple lattice models (on a cubic lattice). One representative class of systems where such a theory should be applicable, is so-called the \emph{multi-Weyl semimetals}. These systems possess linear dispersion only along one component of the momentum, while displaying finite band curvature along the remaining two crystalline directions, see Fig.~\ref{Fig:spectra_multiWeyl}. The power-law dependence of the band dispersion ($n$) is set by the charge $n$ of the corresponding pairs of (anti-)monopole in the momentum space that act as source and sink of Abelian Berry curvature, and in turn also determines the integer topological invariant of the system. Therefore, the present discussion should allow us to pave the path to connect the notion of quantum anomalies with the topological invariant of gapless chiral systems.

Our main achievements are the followings. We show that multi-Weyl semimetals (with $n>1$) generically exhibit non-Abelian anomalies, leading to the non-conservation of isospin density. We also show that both Abelian and non-Abelian anomaly coefficients are solely determined  by the topological invariant ($n$) of the system. Notice that quantum field theories depict three types of chiral anomalies: \emph{Abelian, non-Abelian and gravitational}. While negative longitudinal magnetoresistance, bearing the signature of Abelian anomaly, has been observed in a number of Weyl materials~\cite{Exp1,Exp2,Exp3,Exp4,Exp5,Exp6,Exp7,Exp8}, and some of them also show indirect signature of mixed gauge-gravity anomaly in the thermal transport~\cite{Gooth:2017mbd,Schindler:2018wrd,2016PNASLDS,Manes:2018mth} the presence/signature of non-Abelian anomaly has remained illusive so far. Also the regime where high-energy systems may support non-Abelian anomalous transport, is quite challenging to access experimentally till now \cite{PhysRevD.97.085020,KHARZEEV20161}. By contrast, we here show that multi-Weyl semimetals constitute the ideal platform to capture the signature of non-Abelian anomaly in table-top experiments (from the non-conservation of isospin density), (possibly) standing as the final milestone of anomaly controlled transports. And as a collateral consequence, in this work we liberate the notion of Abelian (or generically any) anomaly from the burden of the Lorentz invariance.

Besides the genuine fundamental importance of our quest, it should also be relevant for real materials, as Weyl points with $n=2$ (known as double-Weyl nodes) can be found in HgCr$_2$Se$_4$~\cite{PhysRevLett.107.186806,PhysRevLett.108.266802} and SrSi$_2$~\cite{Huang1180}, whereas A(MoX)$_3$ (with A=Rb, Tl and X=Te) can accommodate Weyl points with $n=3$ (known as triple-Weyl nodes)~\cite{PhysRevX.7.021019}. Even though at a formal level our conclusions hold for any arbitrary integer value of $n$, crystalline environment forbids realization of symmetry protected Weyl nodes with $n>3$~\cite{Nagaosa}.

We note that a direct approach to construct an effective field theory for multi-Weyl semimetals has been discussed previously in Refs.~\cite{PhysRevB.96.085201, Lepori2018}, where a Lagrangian with an anisotropic energy spectrum and the corresponding anomalous violation of chiral symmetry was computed. We emphasize that this approach is cumbersome and may even be problematic for the following reasons. First of all, we stress that departure from the Lagrangian that is linear in (space-time) four-momenta, changes the structure of fermionic operators and all the anomalies need to be calculated from the scratch. Secondly, the Lagrangian corresponding to anisotropic dispersion obscures the underlying symmetry structure (and may even hinder additional features, about which more in a moment). Finally, previous studies on Lorentz violating theories suggest that certain ambiguities may appear in the formulation that cannot be removed within the effective theory description~\cite{Grushin:2012mt}. To circumvent these pitfalls and address the anomaly structure in multi-Weyl systems in an unambiguous and transparent fashion, we here develop a completely new theoretical approach, highlighted below. For the sake of concreteness, we focus on the minimal model for time-reversal symmetry breaking Weyl semimetals, composed of only a single pair of (anti-)monopole of charge $n$.

Given this motivation, we construct an effective field theory for multi-Weyl systems that is always linear in all momenta, but accompanied by a Lorentz violating perturbation, that ultimately leads to the multi-Weyl spectrum in the low energy limit, see Fig.~\ref{Fig:spectra_multiWeyl}. As we show this formulation has several advantages over the direct approach and improves the analysis in every aspect mentioned above. For instance, from a computational point of view, we do not need to perform additional computations of anomalies, as the theory is always linear. Furthermore, we find that the requisite Lorentz symmetry breaking perturbation yielding the multi-Weyl spectra at low-energy, couples to linearly dispersing chiral fermions as a $SU(2)$ non-Abelian constant gauge field. As a result the anomaly structure is much richer than the ones inferred from previous studies~\cite{PhysRevB.96.085201, Lepori2018}. Namely, in addition to the usual (but generalized) $U(1)$ anomalies, we also unveil non-Abelian $SU(2)$ anomalies for multi-Weyl semimetals when $n>1$. Due to the extensive nature of our study, it is worth pausing at this point to offer an overview of the main results, before delving into the details.

\subsection{Extended Summary}

The minimal effective low-energy model for a multi-Weyl semimetal can be described in terms of two-component chiral (left or right) fermions [see Sec.~\ref{Sec:Lowenergy_MultiWeyl}]. The resulting quasiparticle spectra in the vicinity of each Weyl node scale as $E \sim |p_z|$ and $E \sim p_\perp^n$, where $p_\perp=[p^2_x + p^2_y]^{1/2}$ [see Fig.~\ref{Fig:spectra_multiWeyl}], since in our construction the Weyl nodes are separated along the $z$ direction. Here, $n$ is an integer that determines the (anti-)monopole charge of the Weyl nodes and hence the topological invariant of the system. Therefore, when $n>1$ the energy dispersion in the $xy$ plane displays nontrivial band curvature. In Sec.~\ref{Sec:Lowenergy_MultiWeyl}, we also show that such a nonlinear dispersion for multi-Weyl semimetal can be achieved at low energies by coupling $n$ copies of simple Weyl fermions (with $n=1$, possessing only linear dispersion, see Fig.~\ref{Fig:spectra_multiWeyl}) with a $C_4$ symmetry preserving perturbation ($\Delta$). In the language of effective field theory, such a perturbation breaks the Lorentz invariance and couples with simple Weyl fermions as an $SU(2)$ non-Abelian constant gauge field. Consequently, a multi-Weyl semimetal gets immersed in a constant non-Abelian magnetic field ${\bf B}_3 \sim \Delta^2$. Nonetheless, these two constructions are shown to be equivalent as they yield identical band dispersion at low energies and the topological invariant of the system. But, the later construction allows us to derive an effective field theory for generalized Weyl systems with $n>1$ in terms of simple Weyl fermions, subject to Lorentz symmetry breaking perturbation that enormously simplifies the analysis and keeps the outcomes transparent.

In Sec.~\ref{subsec:onlylatticemodels}, we introduce simple tight-binding models for multi-Weyl semimetals on a cubic lattice. First, we present effective two band models for such systems with $n=1,2$ and $3$, and argue that discrete four-fold rotational ($C_4$) symmetry protects such higher order touchings of Kramers non-degenerate valence and conduction bands at two Weyl nodes. Therefore, multi-Weyl nodes are \emph{symmetry protected}. The resulting band structures are shown in Fig.~\ref{Fig:Bandstructure_2band}. Moreover, we show that multi-Weyl semimetals with $n=2$ and $3$ can be constructed by coupling $n$ copies of the lattice model for simple Weyl semimetal (with $n=1$) by the $C_4$ symmetry preserving perturbation. The resulting band structures of these two systems, shown in Fig.~\ref{Fig:Bandstructure_multiband}, are identical to the ones obtained from their corresponding two-band models [see Fig.~\ref{Fig:Bandstructure_2band}(b) and (c)], but only at low energies.

The topological equivalence between these two constructions for the multi-Weyl systems is then further substantiated from the bulk-boundary correspondence, encoded through the number of Fermi arc surface states, see Sec~\ref{subsec:fermiarc}. Note that a multi-Weyl semimetal, constituted by (anti-)monopole of charge $n$, supports $n$ copies of the Fermi arcs. Indeed we find $n$ copies of the Fermi arcs connecting two Weyl nodes of charge $n$ from the two-band models, see Fig.~\ref{Fig:Fermiarcs_2band}. Furthermore, we also observe $2$ and $3$ copies of the arc states for double and triple Weyl semimetals, respectively, when they are constructed by coupling $2$ and $3$ copies of simple Weyl fermions by a $C_4$ symmetry preserving perturbation, see Fig.~\ref{Fig:Fermiarcs_multiband}.

Upon constructing multi-Weyl semimetals by coupling simple Weyl fermions with a static non-Abelian $SU(2)$ gauge field, we derive the effective field theory of these systems in Sec.~\ref{Sec:QFT_MultiWeyl}. The effective field theory for multi-Weyl semimetals displays both $U(1)$ as well as $SU(2)$ non-Abelian (only for $n>1$) anomalies. To this end, we compute the Ward identities for both covariant and consistent (related by the Bardeen-Zumino polynomials) Abelian and non-Abelian currents. One of our main results is the generalization of anomalous Hall effect for multi-Weyl semimetals, and its non-Abelian variation, respectively captured by, for example, the regular ($\rho_e$) and ``isospin" ($\rho_3$) charge densities, given by     
\begin{equation}~\label{Eq:anomalysummary_intro}
\rho_e = n \; \frac{e^2}{2 \pi^2} \; ({\bf b} \cdot {\bf B}), \:\:
\rho_3= \frac{c(n)}{2 \pi^2} \; ({\bf b} \cdot {\bf B}_3).
\end{equation}
In the above expressions, $e$ is the electric charge, $2|{\bf b}|$ is the separation of left and right Weyl nodes, ${\bf B}$ is the external Abelian magnetic field, ${\bf B}_3=(0,0,\Delta^2)$ is the static non-Abelian magnetic field (present only for $n>1$), $c(n)$ is a coefficient set by the representation of $SU(2)$ Lie group. Specifically, $c(n)=1/2$ and $2$ for $n=2$ and $3$, respectively. We also find that the chiral magnetic effect vanishes for  both vector Abelian and non-Abelian currents (see discussion in Sec.~\ref{sec:4b}, after Eq.~(\ref{eq:abj})).

To test the validity of the field theoretic predictions from Sec.~\ref{Sec:QFT_MultiWeyl}, we first compute Abelian or $U(1)$ charge density ($\rho_e$) in the presence of a static external magnetic field from all the lattice models for multi-Weyl systems, introduced in Sec.~\ref{Sec:LatticeModel}. The methodology is discussed in Sec.~\ref{Sec:QFT_MultiWeyl_Lattice} and the results are displayed in Fig.~\ref{Fig:AbelianAnomaly_Summary}. We find that the field theoretic predictions [see Eq.~(\ref{Eq:anomalysummary_intro})] show excellent agreement with the scaling of the Abelian charge density with the external magnetic field flux, at least when it is small (cyclotron frequency being much smaller than lattice momenta), irrespective of the microscopic details. As a penultimate topic, we compute the non-Abelian or isospin density ($\rho_3$), capturing the signature of non-Abelian anomalies [see Eq.~(\ref{Eq:anomalysummary_intro})], for the multi-Weyl semimetals with $n=2$ and $3$, but only from their four and six band lattice models, respectively. The results are shown in Fig.~\ref{Fig:nonAbelian_summary}, displaying an excellent agreement with the field theoretic predictions.

The topological nature of anomalies in certain cases protects their associated transport, showing universalities even when some symmetries are broken~\cite{Amado:2014mla,Copetti:2016ewq}. Therefore, the microscopical details of different models become irrelevant, as long as the anomalous structure does not differ between them~\cite{Son:2009tf,Neiman:2010zi}. On the other hand, the computation of anomaly induced transport coefficients with standard quantum field theory techniques can be plagued with subtleties and ambiguities~\cite{Jackiw:1999qq,PhysRevB.88.245107,Grushin:2012mt,Rebhan:2009vc,Landsteiner:2013aba,PhysRevB.86.115133,PhysRevX.4.031035}, which have been solved and understood with the help of the holographic techniques~\cite{Gynther:2010ed,Landsteiner:2012kd}. Therefore, we address the imprint of the symmetry breaking parameter ($\Delta$) in various anomaly induced transports from a simple toy model for a (strongly) interacting multi-Weyl semimetal using the holographic techniques, see Sec.~\ref{sec:HOLO}. The particular model we study is consistent with the predictions of the effective field theory, and shows a renormalization of the non-Abelian current in the infrared regime, as expected due to the explicit symmetry breaking introduced by $\Delta$. The main outcome from this section is the survival of the non-Abelian transport at low energies, opening a possibility of observing non-Abelian anomaly and the non-renormalization of the Abelian anomaly-induced transport in multi-Weyl semimetals. Therefore, altogether the current discussion presents a comprehensive study of anomalies in Lorentz symmetry violating multi-Weyl semimetals, which in future can be extended to address similar issues for multi-fold fermions~\cite{2016arXiv160303093B,2017Natur.547..298B}.

\subsection{Outline}

The rest of the paper is organized as follows. In the next section, we discuss the low-energy models for multi-Weyl semimetal and compute it topological invariant. Section~\ref{Sec:LatticeModel} is devoted to the discussion on the lattice models for these systems on a cubic lattice. In this section we also establish the bulk-boundary correspondence by constructing (numerically) the Fermi arc surface states for multi-Weyl semimetal. The effective field theories, capturing the signature of quantum anomalies, for multi-Weyl semimetals are derived in Sec.~\ref{Sec:QFT_MultiWeyl}. The field theoretic predictions from this section are numerically anchored from the representative tight-binding models in Sec.~\ref{Sec:QFT_MultiWeyl_Lattice}. The holographic description and transport coefficients are derived from the gauge-gravity duality in Sec.~\ref{sec:HOLO}. Discussions on our findings and some future directions are highlighted in Sec.~\ref{Sec:Discussion}. Additional technical details are relegated to the appendices.

\section{Multi-Weyl fermions}~\label{Sec:Lowenergy_MultiWeyl}

We begin the discussion by focusing on the effective low-energy models for multi-Weyl systems, constituted by a pair of (anti-)monopole of charge $n$~\cite{PhysRevB.96.085201,PhysRevB.96.085130,PhysRevB.99.115109,PhysRevB.99.075153,2019arXiv190106716M,PhysRevLett.107.186806,PhysRevLett.108.266802,Huang1180,PhysRevX.7.021019,Nagaosa,PhysRevB.92.125141,PhysRevB.94.195144,RevModPhys.90.015001}, where $n$ is an integer. The Hamiltonian operator describing such system takes the form   
\begin{equation}~\label{eq:hmulti}
H^\pm_{n} = \alpha_{n} p^n_{_\bot} \left[ \cos \left( n \phi_{p} \right) \tau_{x} +\sin \left( n \phi_{p} \right) \tau_{y} \right] \pm v p_z \tau_{z},
\end{equation}
where $p_{_\perp} = (p^2_x + p^2_y)^{1/2}$ and $\pm$ correspond to two valleys, respectively acting as the monopole (source) and antimonopole (sink) of Abelian Berry curvature. Around these two points low-energy excitations are described in terms of left and right chiral fermions, respectively. Momentum ${\bf p}$ is measured from the Weyl node. The set of Pauli matrices $\boldsymbol{\tau}= \left( \tau_x, \tau_y, \tau_z \right)$ operate on the pseudospin indices. The energy spectra in the close vicinity of the Weyl nodes take the form $\pm \epsilon_{\mathbf p}$, where $\pm$ respectively correspond to the conduction and valence bands, and 
\begin{equation}\label{eq:espectrum}
\epsilon_{\mathbf{p}} =  \sqrt{ \alpha^2_{n} p^{2 n}_{_\bot} + v^2 p^2_z}. 
\end{equation} 
Respectively, for $n=1$ and $2$ the parameter $\alpha_n$ bears the dimension of velocity and inverse mass, while $v$ is the Fermi velocity in the $z$ direction. The energy dispersions along various high symmetry directions for $n=1,2$ and $3$ are shown in Fig.~\ref{Fig:spectra_multiWeyl}.

The topological invariant of Weyl systems is given by the integer (anti-)monopole charge, which can be computed in the following way. For concreteness, we now focus near one valley, hosting left chiral fermions and introduce the following coordinate system    
\begin{eqnarray}
\left( p_x, p_y, p_z \right) = \left( p_{_\perp} \cos \phi, p_{_\perp} \sin \phi, \frac{\epsilon_{\mathbf p}}{v} \cos \theta \right),
\end{eqnarray}
where $p_{_\perp}=\left( \epsilon_{\mathbf p} \sin \theta/\alpha_n \right)^{1/n}$. The Berry curvature of the conduction band then takes the form \cite{Dantas:2018udo}
\begin{align}
\mathbf{\Omega}_{\mathbf{p}} {}= \frac{n^2 \; \alpha^2_n }{2 \; \epsilon^2_{\bf p} } \left( \frac{\epsilon_{\bf p} \; \sin \theta}{\alpha_n} \right)^{\frac{2(n-1)}{n}} h_1 \; \hat{\mathbf{\epsilon}},
\end{align}
where $\hat{\mathbf{\epsilon}} $ is the unit-norm radial vector and 
\begin{equation}
h_1= \frac{1}{v} \; \left[\cos^2 \theta + \frac{v^2}{n^2 \epsilon^2_{\mathbf{p}}} \left(\frac{\epsilon_{\mathbf{p}} \sin \theta }{ \alpha_n }\right)^{2/n}\right]^{1/2}.
\end{equation}
The integer monopole charge can then be obtained by integrating the Berry curvature over a unit sphere ($\Sigma$), defined by $\epsilon_{\bf p}=1$, around the Weyl node, yielding  
\begin{equation}~\label{eq:topological}
\frac{1}{2 \pi} \oint_{\Sigma} \mathbf{\Omega}_{\mathbf{p}} \cdot d\mathbf{S}=n.
\end{equation}

\begin{figure*}[t!]
\subfigure[]{
\includegraphics[width=0.32\linewidth]{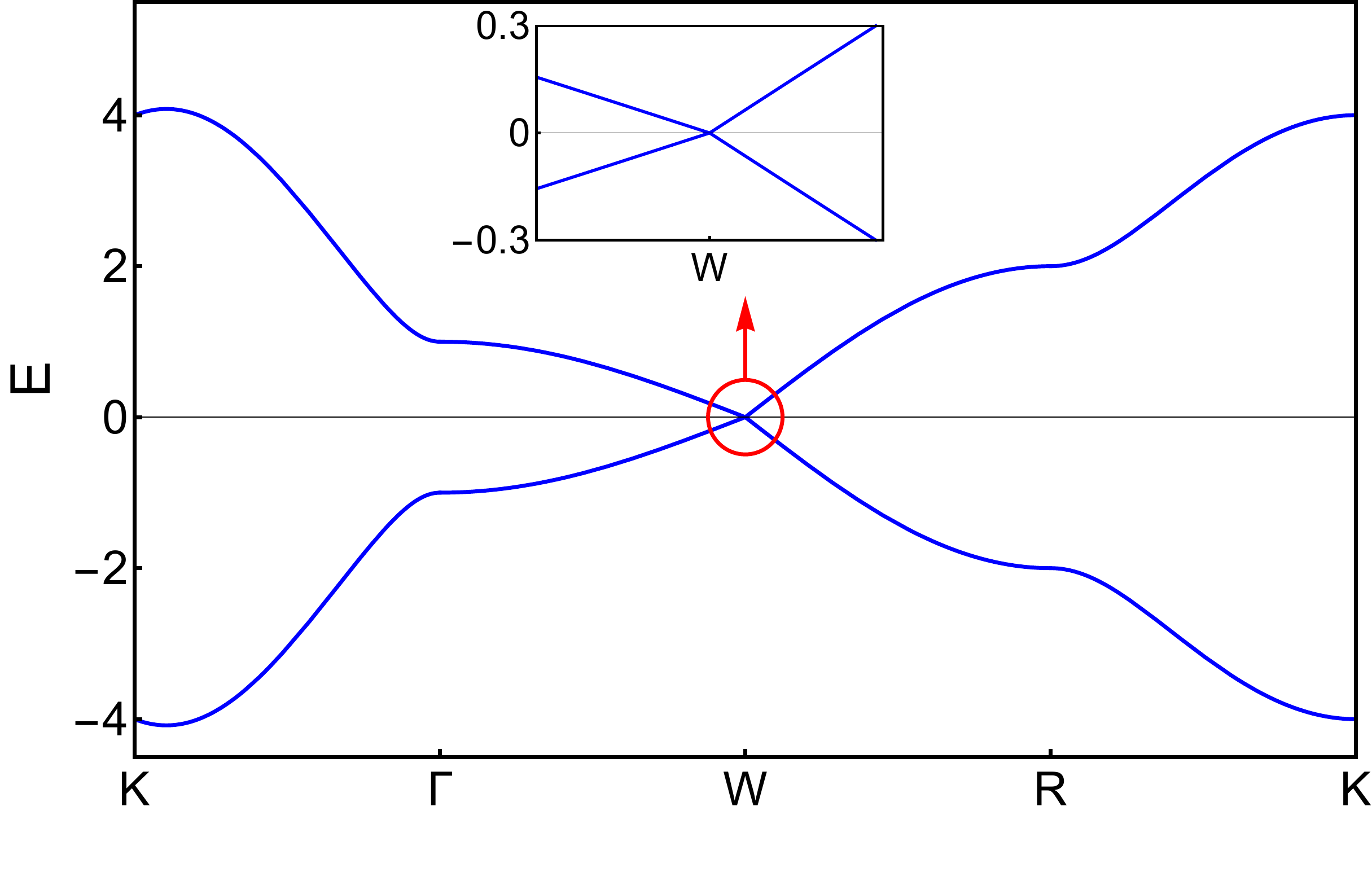}
}%
\subfigure[]{
\includegraphics[width=0.32\linewidth]{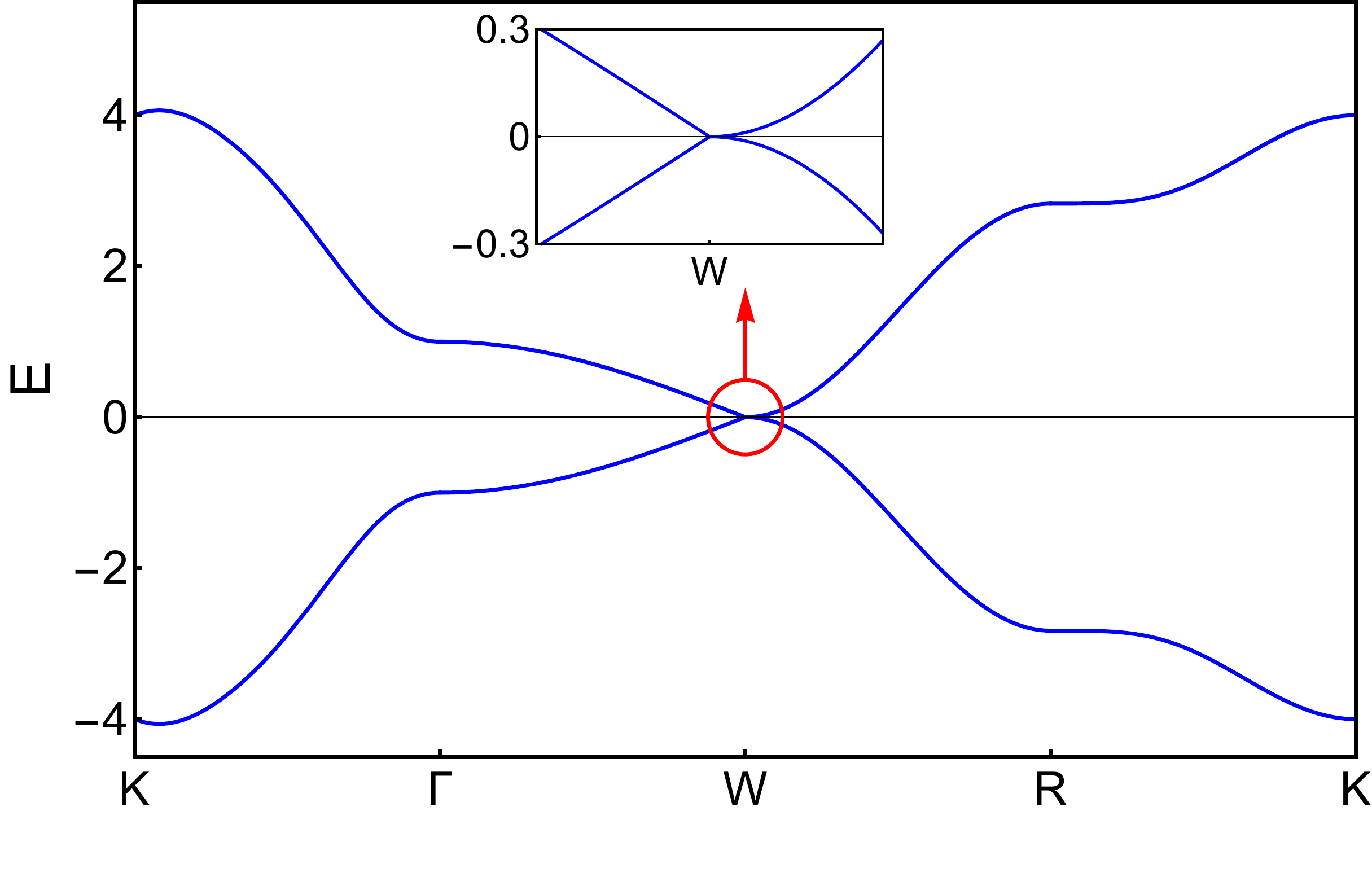}
}%
\subfigure[]{
\includegraphics[width=0.32\linewidth]{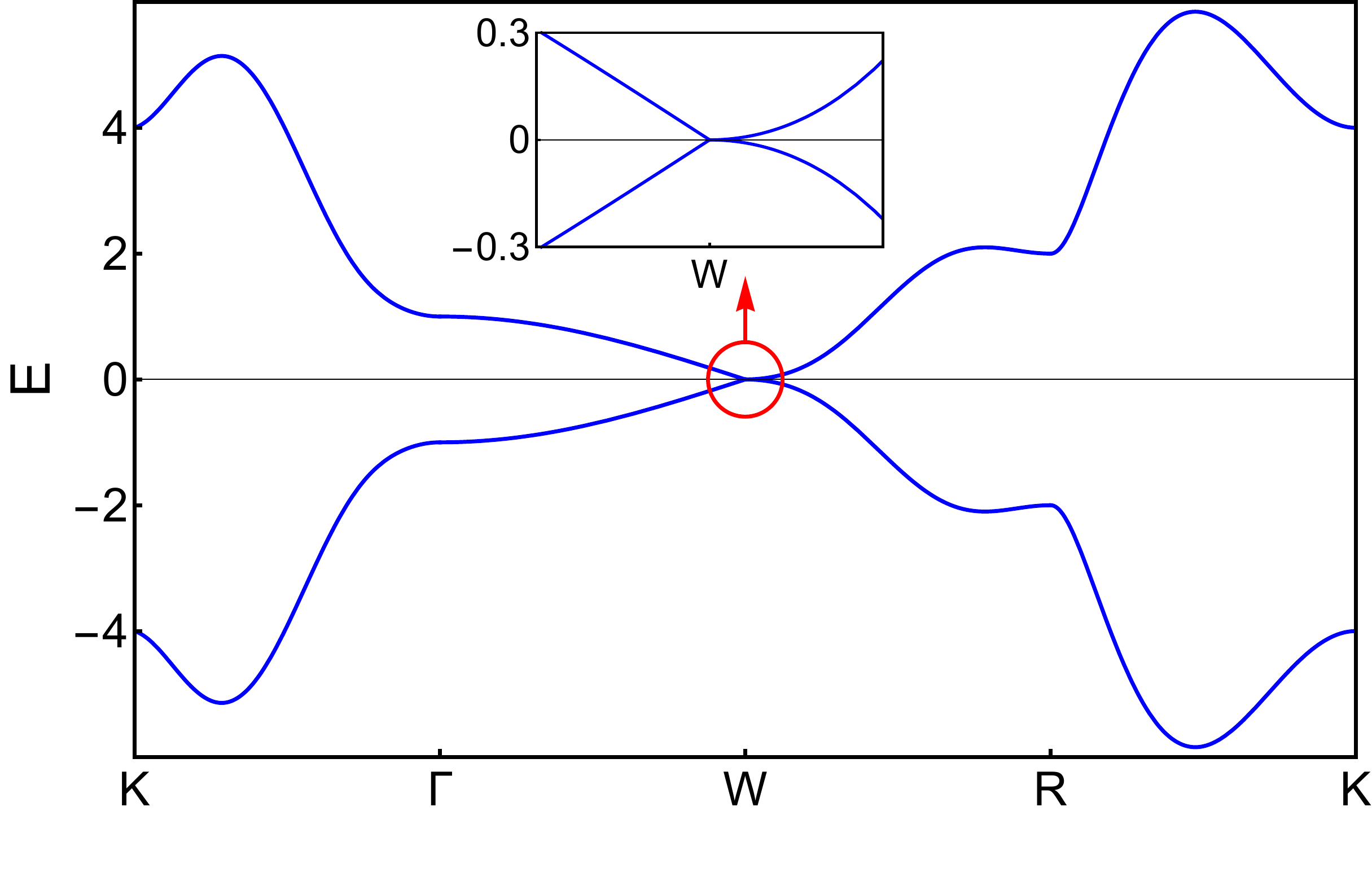}
}%
\caption{ Band structure of (a) simple ($n=1$), (b) double ($n=2$) and (c) triple ($n=3$) Weyl semimetals from the corresponding effective two band models, see Eqs.~(\ref{eq:Weyl_2bandcriptic})-(\ref{eq:Ny_twoband}). We set $t=t_0=t_z=1$ in the tight-binding model, and the lattice constant $a=1$. We here take the path ${\rm K} \rightarrow \Gamma \rightarrow {\rm W} \rightarrow {\rm R} \rightarrow {\rm K}$, where ${\rm K}$, $\Gamma$, ${\rm W}$ and ${\rm R}$ stand for $(\pi,\pi,\pi/2)$, $(0,0,0)$, $(0,0,\pi/2)$ and $(\pi,0,\pi/2)$, respectively. The energy ($E$) is measured in units of $t$. Note that Weyl nodes are located at $(0,0,\pm \pi/2)$, around which dispersion always scales linearly along the $z$ directions, while it respectively display linear, quadratic and cubic dispersion in the $(k_x, k_y)$ plane (see insets). Due to the absence of particle-hole asymmetry, the Weyl nodes are always pinned at zero energy. Realizations of double and triple Weyl systems respectively from four and six band models are shown in Fig.~\ref{Fig:Bandstructure_multiband}.    
}~\label{Fig:Bandstructure_2band}
\end{figure*}

Even though the low energy model for multi-Weyl semimetals correctly captures the topological invariant of the system, one can construct the Weyl models with $n>1$ by coupling $n$ copies (hereafter referred `flavor') of simple Weyl fermions in the following way. This construction follows the spirit of realizing higher order band touchings in multi layer graphenelike systems by introducing interlayer tunneling~\cite{RevModPhys.81.109,PhysRevB.88.075415,Huang1180}. In addition, this construction opens an efficient route to arrive at the effective field theoretic description for Weyl systems with $n>1$ [see Sec.~\ref{Sec:QFT_MultiWeyl}]. We focus near the left chiral valley and introduce the following Hamiltonian operator
\begin{eqnarray}~\label{Eq:genWeyl_coupled}
H^{\rm coup}_n &=& \left[ v_{_\perp} \left( p_x \tau_x + p_y \tau_y \right) + v p_z \tau_z \right] \otimes \mathbb{1} _{n\times n} \nonumber \\
&+& \Delta \; ( \tau_x \otimes s^n_x + \tau_y \otimes s^n_y ),
\end{eqnarray} 
where $\mathbb{1} _{n\times n}$, $s^n_x$ and $s^n_y$ operate on the flavor index, while ${\boldsymbol \tau}$ operate on pseudospin index. Note that the first term in $H^{\rm coup}_n$ corresponds to $n$ decoupled flavors of simple Weyl fermions, while the term proportional to $\Delta$ introduces nontrivial coupling between them~\footnote{Note that the form of such inter-flavor coupling is not unique. One can choose it to be $\Delta \; ( \tau_y \otimes s^n_x - \tau_x \otimes s^n_y )$, which leaves all the physical outcomes unchanged.}. For any integer $n$, $\mathbb{1} _{n\times n}$ is $n$-dimensional identity matrix, ${\boldsymbol s}^{n}$ are the generators of the spin-$(n-1)/2$ representation of $SU(2)$. In particular for $n=2$, ${\boldsymbol s}^{n}={\boldsymbol \sigma}/2$, where ${\boldsymbol \sigma}$ are the Pauli matrices, while for $n=3$
\begin{equation}
s^{n}_x=\frac{\lambda_1+ \lambda_6}{\sqrt{2}}, \;
s^{n}_y=\frac{\lambda_2+ \lambda_7 }{\sqrt{2}}, \;
s^{n}_z=\frac{\lambda_3 + \sqrt{3} \lambda_8 }{2}, \nonumber 
\end{equation} 
and ${\boldsymbol \lambda}$ are the Gell-Mann matrices~\cite{PhysRev.125.1067}. In principle, one can generalize this construction for arbitrary integer value of $n$. But, in crystalline environment only Weyl nodes with $n \leq 3$ are symmetry protected. So, we here focus on Weyl systems with $n=1,2$ and $3$. The derivation of the low-energy Hamiltonian [see Eq.~(\ref{eq:hmulti})] for multi-Weyl semimetals with $n=2$ and $3$, starting from the above coupled models are shown in Appendix~\ref{appendix:lowenergymodel}.

For $n=2$ the energy spectra are composed of four branches, given by $\pm \epsilon^q_{\bf p}$, where 
\begin{equation}~\label{Eq:SpectraDW_4band}
\epsilon^{q}_{\bf p}= \left[ \left(\sqrt{\frac{\Delta^2}{4} + v^2_{_\perp} p_\perp^2} - (-1)^q \: \frac{\Delta}{2} \right)^2 + v^2 p_z^2 \right]^{1/2},
\end{equation}
for $q=0$ and $1$, and $\pm$ correspond to the conduction and valence bands, respectively. Note that only the $q=0$ branch displays band touching at ${\bf p}=0$, while the $q=1$ branch is fully gapped for any $\Delta \neq 0$. Expanding $\epsilon^0_{\bf p}$ for large $\Delta$ and small $p_{_\perp}$, we obtain 
\begin{equation}
\epsilon^0_{\bf p}= \left[ \frac{v^4_{_\perp} p^4_{_\perp}}{\Delta^2} + v^2 p^2_z + {\mathcal O} \left(\frac{ v^6_{_\perp}p_{_\perp}^6}{\Delta^4}\right) \right]^{1/2},
\end{equation}
which agrees with the expression from Eq.~(\ref{eq:espectrum}) to the order $p^4_{_\perp}$, with $\alpha_2=v^2_{_\perp}/\Delta$, bearing the dimension of inverse mass. Shortly, we show that the pair of split off bands are topologically trivial, while the band touching point within the $q=0$ sector act as a monopole of charge $n=2$. On the other hand, for $n=3$, the energy spectra are composed of six branches $\pm \epsilon^{q}_{\bf p}$ for $q=0,1,2$, where 
\begin{eqnarray}
&& \epsilon^{q}_{\bf p} = \bigg[ v^2_{_\perp} p^2 _{_\perp} + v^2 p^2_{z} + \frac {2 \Delta^2} {3} + \frac {2 \Delta} {3} \sqrt{6 v^2_{_\perp} p^2_{_{\perp}}+\Delta^2} \times  \nonumber \\
&&  \cos\left( \frac {1} {3}\cos^{-1} \left[\frac{9 v^2_{_\perp} p^2 _{_\perp}\Delta - 2\Delta^3}{2 (6 v^2_{_\perp} p^2_{ {_\perp}} + \Delta^2)^{3/2}}\right]-\frac{2 \pi (2-q)}{3}\right) \bigg]^{\frac{1}{2}}. \nonumber \\
\end{eqnarray}
Note that only the $q=0$ branch displays band touching at ${\bf p}=0$, which acts as monopole of charge $n=3$, while the remaining four bands are completely gapped and topologically trivial. Expanding $\epsilon^0_{\bf p}$ for large $\Delta$ and small $p_{_\perp}$, we obtain 
\begin{equation}
\epsilon^0_{\bf p}= \left[ \frac{v^6_{_\perp}p^6_{_\perp}}{\Delta^4} + v^2 p^2_z + {\mathcal O} \left( \frac{v^8_{_\perp}p^8_{_\perp}}{\Delta^6} \right) \right]^{1/2},
\end{equation} 
which agrees with Eq.~(\ref{eq:espectrum}) to the order $p^6_{_\perp}$, with $\alpha_3=v^3_{_\perp}/\Delta^2$. The above construction of generating multi-Weyl systems by coupling simple-Weyl fermions is, however, not an artifact of low-enegry approximation. In Sec.~\ref{Sec:LatticeModel}, we show that such construction is operative even when we start from the lattice regularized models.

Finally, we compute the Berry curvature for each band for multi-Weyl systems with $n>1$, obtained by coupling $n$ flavors of simple Weyl fermions. For $n=2$, we perform this exercise analytically, by introducing the nonorthogonal curvilinear coordinate system 
\begin{equation}
p_x= p_{_\perp} \cos{\phi}, \: p_y= p_{_\perp} \sin{\phi}, \: p_z= \epsilon^0_{\bf p} \cos{\theta}, 
\end{equation}
where now $p^2_{_\perp}=\epsilon^0_{\bf p} \sin \theta \; (\epsilon^0_{\bf p} \sin \theta +\Delta)$, and $\epsilon^0_{\bf p}$ is displayed in Eq.~(\ref{Eq:SpectraDW_4band}). In this coordinate system, the Berry curvature for the lower conduction band takes a compact form 
\begin{align}
\mathbf{\Omega} &= -\frac{\Delta \sin{2 \theta}}{ \epsilon^0_{\bf p} \left( \Delta+2 \epsilon^0_{\bf p} \sin{\theta}\right)^3} \; \mathbf{e}_{\theta} \\
&+  \frac{2 \Delta \epsilon^0_{\bf p} + 2 \sin \theta \left[ \Delta^2 + 2 \epsilon^0_{\bf p} \Delta \sin \theta + 2 (\epsilon^0_{\bf p})^2 \sin^2 \theta \right]}{\epsilon^0_{\bf p}\left(\Delta+ 2 \epsilon^0_{\bf p} \sin{\theta}\right)^3} \; \mathbf{e}_{\epsilon}, \nonumber 
\end{align}
where $\mathbf{e}_{\epsilon}=\partial \mathbf{r}/\partial \epsilon^0_{\bf p}$ and $\mathbf{e}_{\theta}=\partial \mathbf{r}/\partial \theta$ are the covariant basis vectors. From the above expression for the Berry curvature, we can immediately compute the integer charge assocaited with the band touching point from Eq.~(\ref{eq:topological}), yielding $n=2$ for any $\Delta \neq 0$. The expression for the Berry curvature for the gapped valence and conduction bands are quite lengthy and not very instructive. However, when we integrate the Berry curvature over a closed surface [see Eq.~(\ref{eq:topological})], it yields a trivial answer. Therefore, in the four-band construction for double Weyl fermions, only the two bands touching each other are topologically nontrivial. For triple Weyl fermions four gapped bands are topologically trivial, while the monopole charge of the band touching points, where valence and conduction bands meet is $n=3$.


\section{Lattice model, Bulk-boundary correspondence and Fermi arcs}~\label{Sec:LatticeModel}

In this section, we introduce effective tight-binding models on cubic lattice yielding multi-Weyl semimetals, possessing only two Weyl nodes. Subsequently, we establish the bulk-boundary correspondence for these systems by computing the Fermi arc surface states. These analyses substantiate our discussion from the last section. In addition, we also subscribe to these lattice models to test the predictions from the effective field theory for multi-Weyl semimetals [see Sec.~\ref{Sec:QFT_MultiWeyl}], discussed in Sec.~\ref{Sec:QFT_MultiWeyl_Lattice}.

\begin{figure}[t!]
\subfigure[]{
\includegraphics[width=0.95\linewidth]{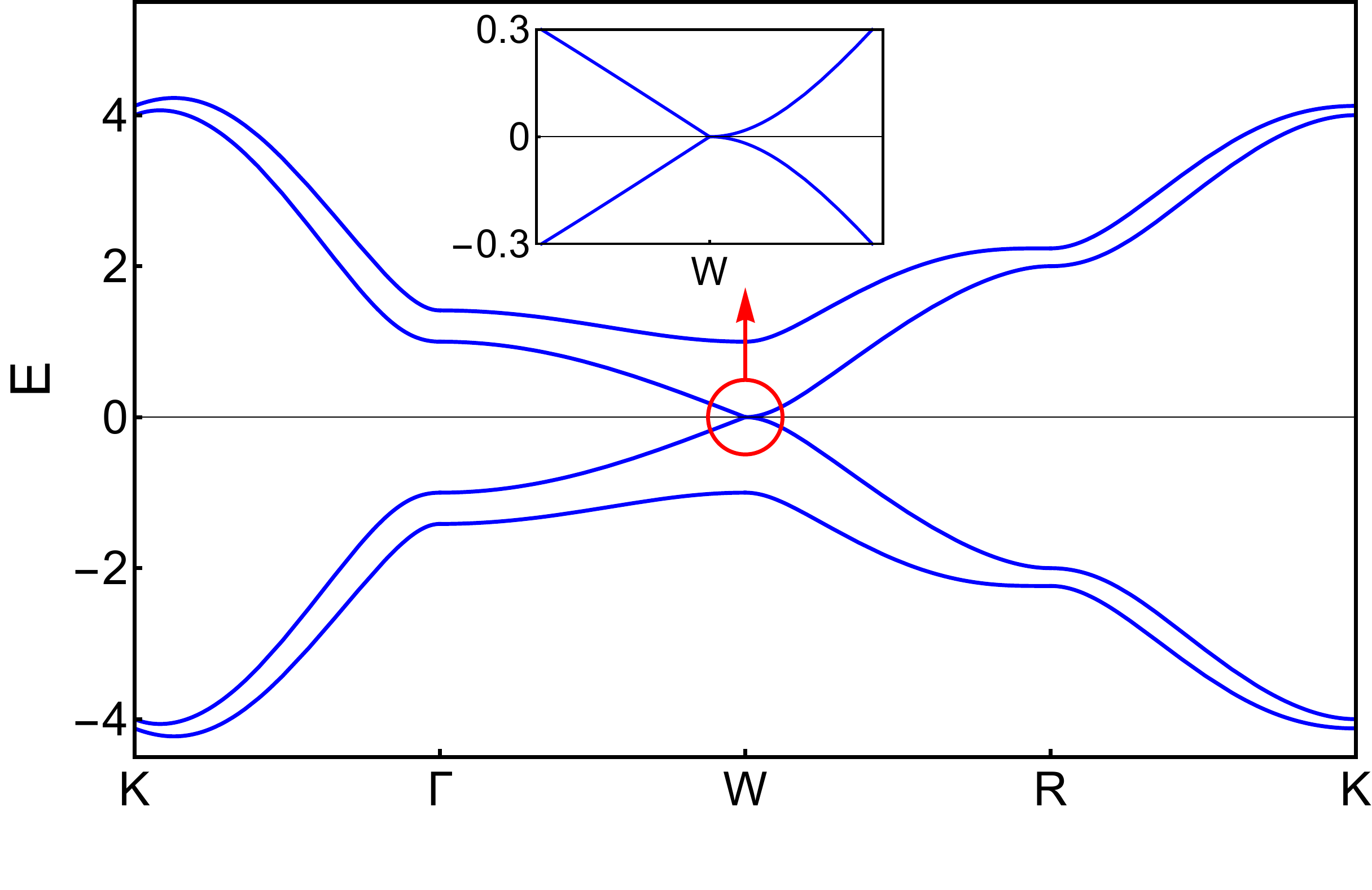}
}
\subfigure[]{
\includegraphics[width=0.95\linewidth]{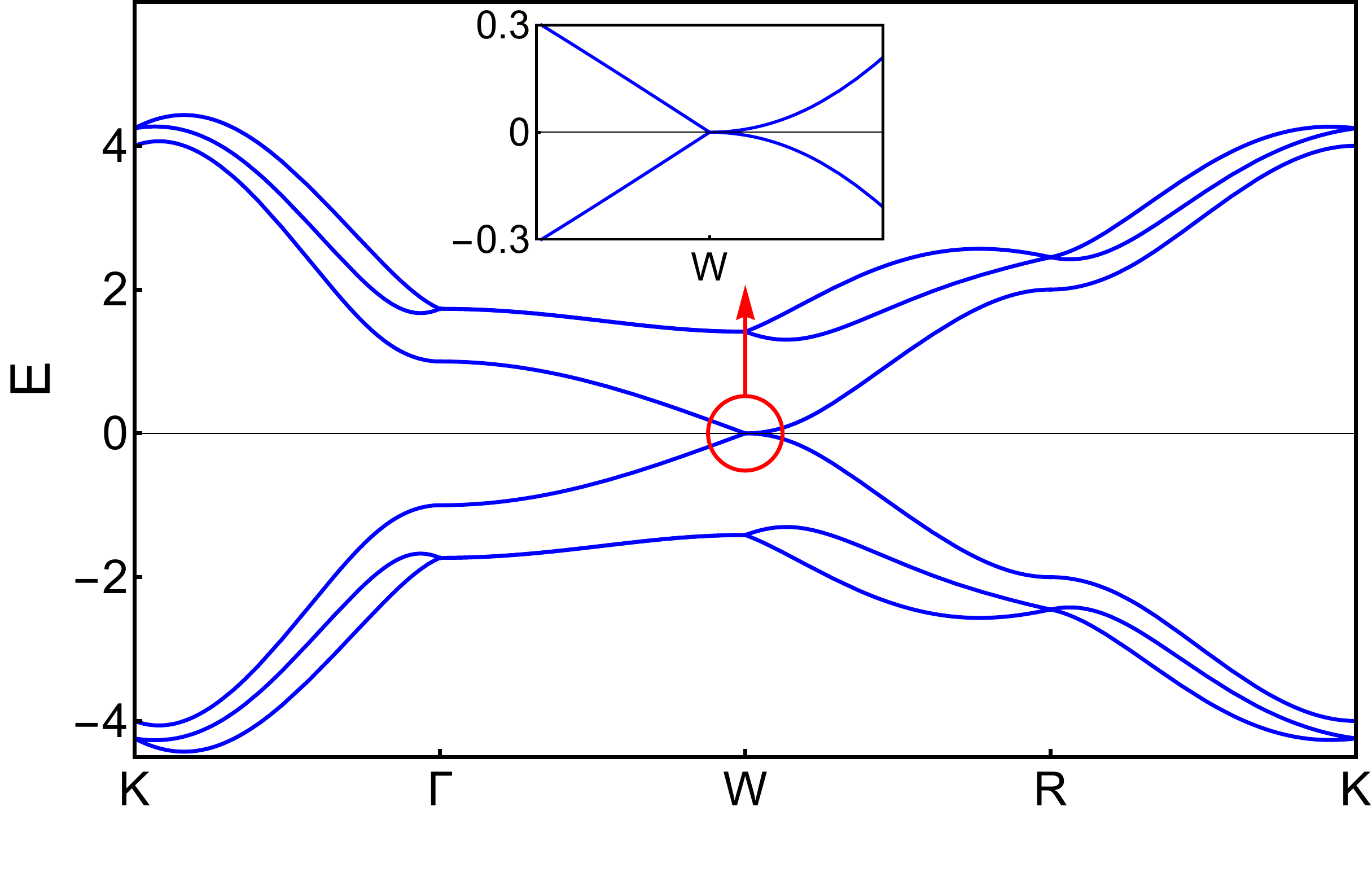}
}
\caption{ Band structure of (a) double and (b) triple Weyl semimetals from four and six band models, respectively [see Eqs.~(\ref{Eq:simpleweyl_TBexplicit}) and (\ref{Eq:genWeyl_coupled_lattice})]. For numerical diagonalization we set $t=t_0=t_z=\Delta=1$ in the tight-binding models, and the lattice spacing $a=1$. Note that these models also display quadratic [see panel (a)] and cubic [see panel (b)] dispersion in the ($k_x,k_y$) plane around the Weyl nodes located at $W=(0,0,\pm \pi/2)$ [compare with subfigures (b) and (c) of Fig.~\ref{Fig:Bandstructure_2band}]. The dispersion always scales linearly with $k_z$.     
}~\label{Fig:Bandstructure_multiband}
\end{figure}

\begin{figure*}[t!]
\subfigure[]{
\includegraphics[width=0.29\linewidth]{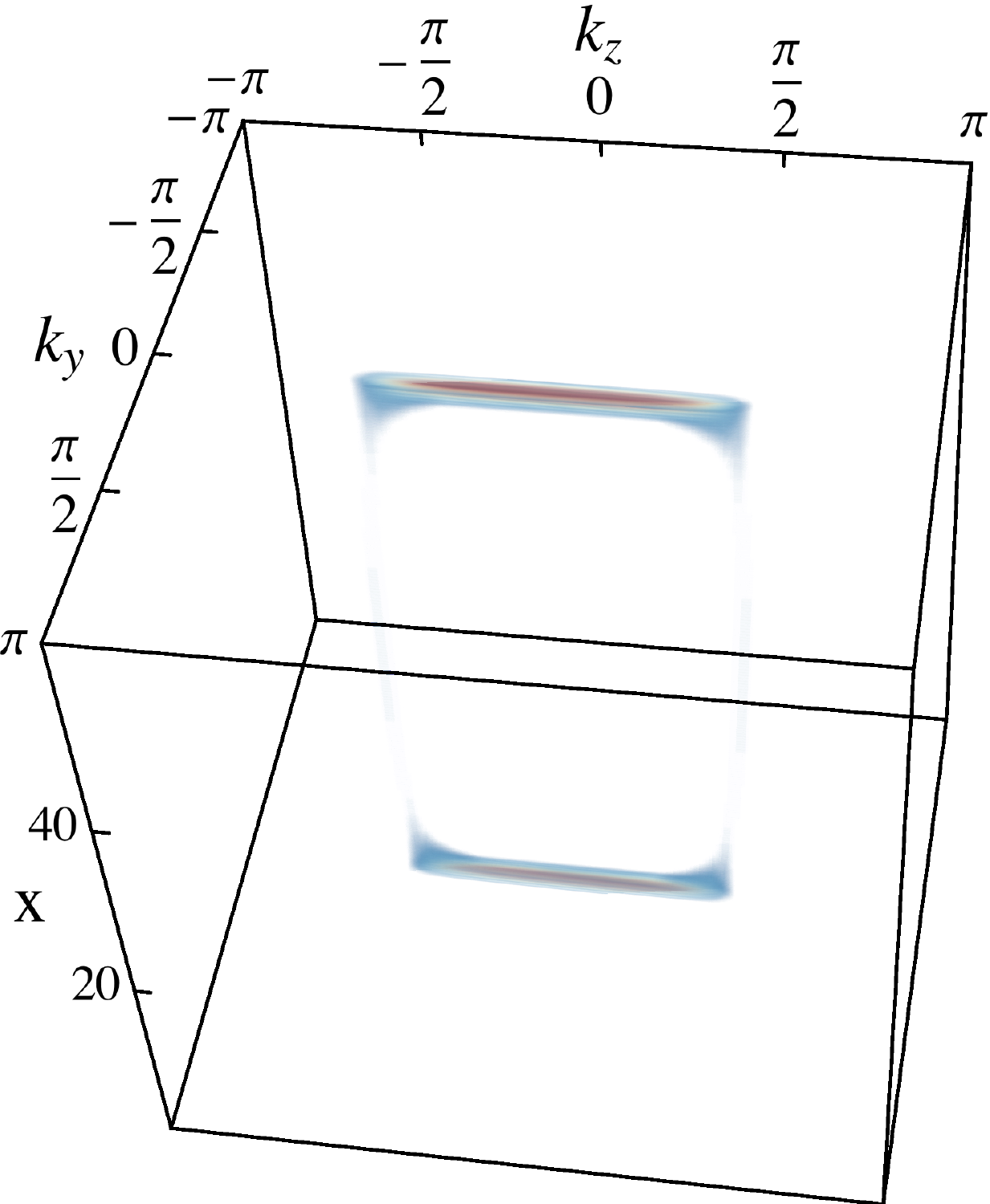}
}%
\subfigure[]{
\includegraphics[width=0.29\linewidth]{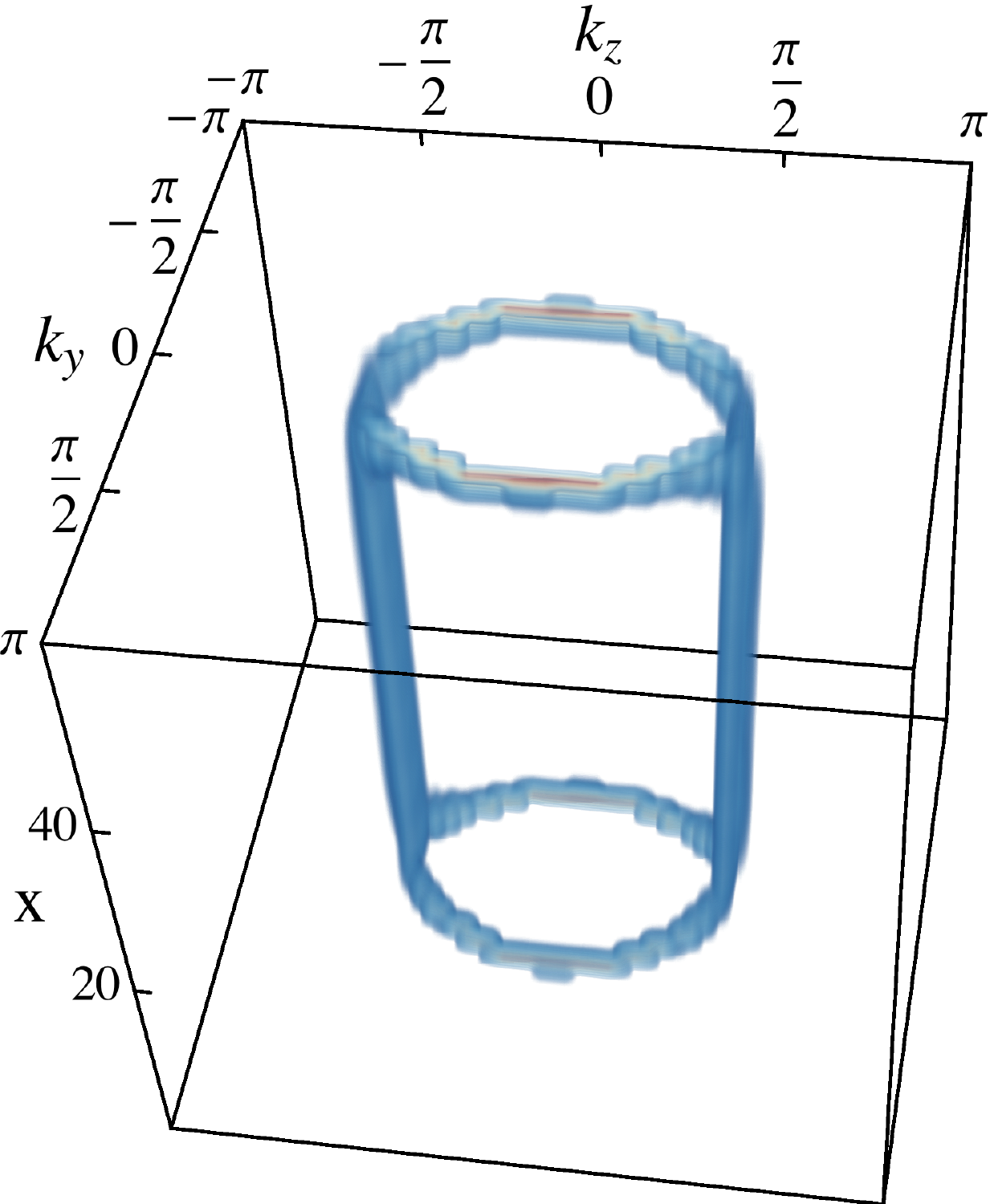}
}%
\subfigure[]{
\includegraphics[width=0.29\linewidth]{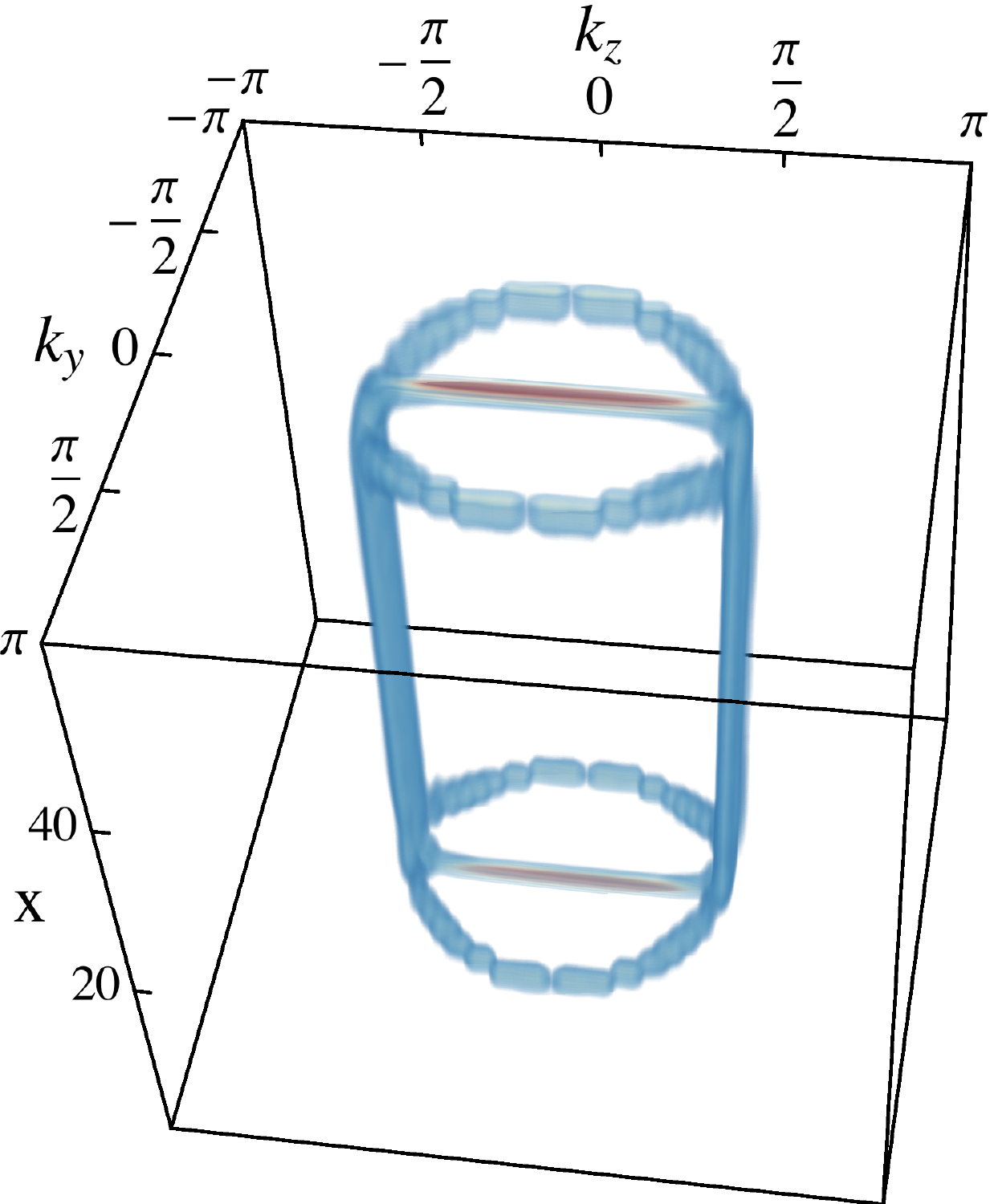}
}%
\includegraphics[width=0.075\linewidth]{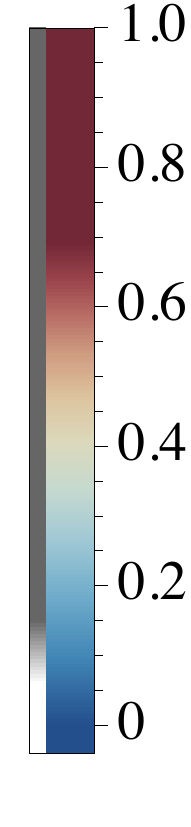}
\caption{Topologically protected Fermi arc surface states for (a) simple ($n=1$), (b) double ($n=2$) and (c) triple ($n=3$) Weyl semimetals, obtained from their two band models [see Eqs.~(\ref{eq:Weyl_2bandcriptic})-(\ref{eq:Ny_twoband})]. For numerical diagonalization we implement a mixed momentum (along $k_y$ and $k_z$) and real space or Wannier (along $x$) representation. The linear dimensionality of the system along $x$ is $L=60$ [see Sec.~\ref{subsec:fermiarc} for details]. Consequently, the Fermi arc surface states are localized on the top and bottom surfaces. We set $t=t_z=t_0=1$. We here display the square of the amplitude of the low-energy states within an energy window $\Delta E=0.1$ for $n=1$, $\Delta E=0.08$ for $n=2$ and $\Delta E=0.05$ for $n=3$ around zero energy. The number of Fermi arcs is equal to $n$, anchoring the bulk-boundary correspondence for multi-Weyl semimetals. In addition, the Fermi arcs from opposite surfaces (top and bottom) get connected through the bulk Weyl points [acting as defects, namely monopole and antimonopole, in the momentum space], where the band gap vanishes. 
}~\label{Fig:Fermiarcs_2band}
\end{figure*}

      \subsection{Lattice models}~\label{subsec:onlylatticemodels}

The lattice model for general Weyl fermions can compactly be written as 
\begin{equation}~\label{eq:Weyl_2bandcriptic}
H_{\rm Weyl}= \sum_{\bf k} \Psi^\dagger_{\bf k} \; \left[ {\bf N} ({\bf k}) \cdot {\boldsymbol \tau}  \right] \; \Psi_{\bf k}, 
\end{equation}
where $\Psi^\top_{\bf k} = \left( c_{{\bf k}, \uparrow}, c_{{\bf k}, \downarrow} \right)$ is a two-component spinor, and $c_{{\bf k}, \tau}$ is the fermion annihilation operator with momentum ${\bf k}$ and pseudospin projection $\tau=\uparrow, \downarrow$. The effective two-band theory emerging from the above tight-binding model give rise to left and right chiral Weyl fermions respectively near $(0,0,\pm \frac{\pi}{2a})$, if we choose 
\begin{equation}
N_3 ({\bf k})= t_z \cz + t_0 [2-\cx - \cy].
\end{equation}  
For convenience, we set the lattice spacing $a$ to be unity. Then simple, double and triple Weyl fermions are realized when we take~\cite{PhysRevB.95.201102,PhysRevB.93.201302,PhysRevX.8.031076}  
\begin{equation}~\label{eq:Nx_twoband}
 N_x({\bf k})= t \:
\begin{cases}
 \Sx & \text{for}\:\: n=1,\\
 \Cx -\Cy & \text{for}\:\: n=2,\\
 \Sx \left[ 3 \Cy-\Cx-2 \right] & \text{for}\:\: n=3,
\end{cases} 
\end{equation}
and
\begin{equation}~\label{eq:Ny_twoband} 
 N_y({\bf k})= t \:
\begin{cases}
 \Sy & \text{for}\:\: n=1,\\
 \Sx \Sy & \text{for}\:\: n=2,\\
 \Sy \left[ 3 \Cx-\Cy-2 \right] & \text{for}\:\: n=3. 
\end{cases}
\end{equation}
The resulting band structures for $n=1,2$ and $3$ are shown in Fig.~\ref{Fig:Bandstructure_2band}. Notice that the above tight-binding models produce only a pair of Weyl nodes at $(0,0,\pm \pi/2)$, around which the effective low-energy models assume the form announced in Sec.~\ref{Sec:Lowenergy_MultiWeyl} [see Eq.~(\ref{eq:hmulti})].

The band touching points in multi-Weyl semimetals are protected by the four-four or $C_4$ rotation about the $z$ axis, a bonafide symmetry operation of $D_{4d}$ point group. Under such a $C_4$ rotation $(k_x,k_y,k_z) \to (-k_y,k_x,k_z)$. When such rotation in the momentum space is accompanied by a rotation by an angle $\theta^n_{\rm SP}=n \frac{\pi}{2}$ in the pseudo spin space, captured by the unitary operator ${\mathcal R}_{\rm SP}(n \frac{\pi}{2})= \exp[i \theta^n_{\rm SP} \tau_z]$, the Hamiltonian operator for $n=1$ and $2$ remains completely invariant. The situation for $n=3$ is slightly more subtle, as $N_{x,y}({\bf k}) \to -N_{x,y}({\bf k})$. Nonetheless, monopole and anti-monopole maps onto themselves under such $C_4$ rotations, leaving the triple-Weyl points symmetry protected. On the other hand, if we take $N_{x}({\bf k}) \leftrightarrow N_y ({\bf k})$ for $n=3$, all the outcomes remain unchanged, but the corresponding Hamiltonian operator remains completely invariant under the $C_4$ rotation. Hence, multi-Weyl points are symmetry protected in a system possessing a $D_{4d}$ symmetry.

The multi-Weyl semimetals with $n>1$ can also be realized by properly coupling $n$ copies of simple Weyl semimetals. We discussed this construction from the continuum or low-energy models in Sec.~\ref{Sec:Lowenergy_MultiWeyl}. We now test the validity of such a construction, starting from the lattice models for $n=1$ Weyl fermions, given by 
\begin{eqnarray}~\label{Eq:simpleweyl_TBexplicit}
H_{\rm SW} &=& t \left[ \Sx \tau_x +\Sy \tau_y \right] \\
&+& \left[ t_z \Cz + t_0 (2-\Cx-\Cy) \right] \tau_z. \nonumber
\end{eqnarray}  
Following Eq.~(\ref{Eq:genWeyl_coupled}), we construct the lattice model for multi-Weyl semimetals by coupling $n$ copies simple Weyl semimetals according to 
\begin{eqnarray}~\label{Eq:genWeyl_coupled_lattice}
H^{\rm coup}_{n, \rm latt} =  H_{\rm SW} \otimes \mathbb{1} _{n\times n} 
+ \Delta \; ( \tau_x \otimes s^n_x + \tau_y \otimes s^n_y ).
\end{eqnarray} 
The notation is the same as in Sec.~\ref{Sec:Lowenergy_MultiWeyl}. The resulting band structure for $n=2$ and $3$ are shown in Fig.~\ref{Fig:Bandstructure_multiband}. Respectively for $n=2$ and $3$, two and four bands are completely gapped, while the remaining two bands touch each other at $(0,0,\pm \pi/2)$. The energy dispersions around these points are respectively quadratic and cubic with the in-plane components of momenta, but always scales linearly with its $z$ component.

      \subsection{Fermi arcs}~\label{subsec:fermiarc}

\begin{figure}[t!]
\subfigure[]{
\includegraphics[width=0.48\linewidth]{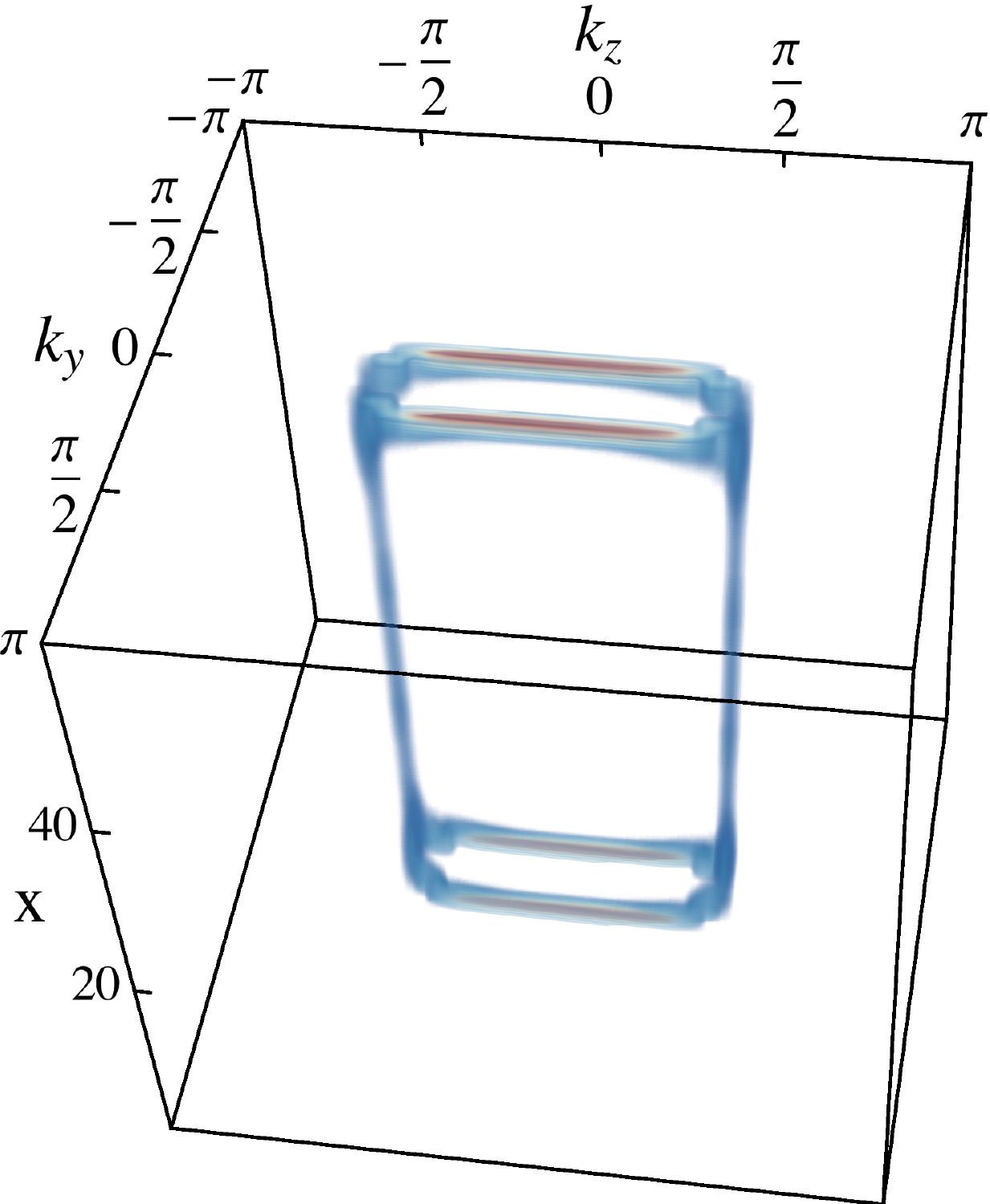}
}%
\subfigure[]{
\includegraphics[width=0.48\linewidth]{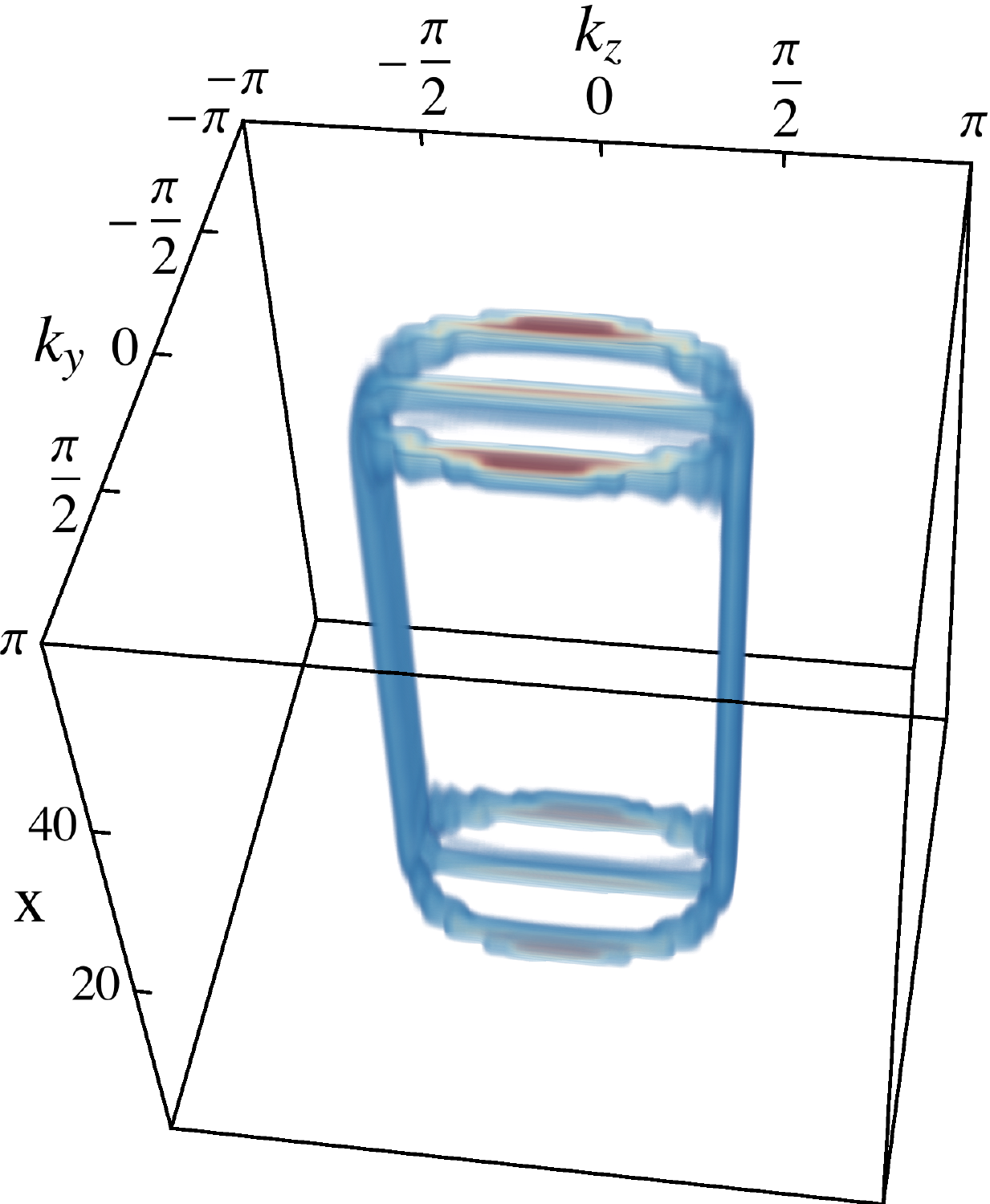}
}
\caption{ Fermi arc surface states for (a) double ($n=2$) and (b) triple ($n=3$) Weyl semimetals, obtained respectively from four and six band models [see Eqs.~(\ref{Eq:simpleweyl_TBexplicit}) and ~(\ref{Eq:genWeyl_coupled_lattice})]. For numerical diagonalization we follow the same approach, mentioned in the caption of Fig.~\ref{Fig:Fermiarcs_2band}, and set $t=t_z=t_0=\Delta=1$, $a=1$. We here display the square of the amplitude of the low-energy states within an energy window $\Delta E=0.05$ for $n=2$ and $\Delta E=0.04$ for $n=3$. Note that double and triple Weyl semimetals respectively support two and three copies of Fermi arc surface states, as we obtained from their two band representations [see (b) and (c) of Fig.~\ref{Fig:Fermiarcs_2band}]. This observation establishes the topological equivalence among these models and the bulk-boundary correspondence in these gapless topological systems. The color scale is the same as in Fig.~\ref{Fig:Fermiarcs_2band}. 
}~\label{Fig:Fermiarcs_multiband}
\end{figure}

Previously in Sec.~\ref{Sec:Lowenergy_MultiWeyl}, we showed that the four and six band models respectively for double and triple Weyl fermions [see Eq.~(\ref{Eq:genWeyl_coupled})] and their low-energy description in terms of the two band models [see Eq.~(\ref{eq:hmulti})] yield identical topological invariant (the monopole charge). The monopole charge determines the integer topological invariant of the system that in turn also dictates the number of topologically protected Fermi arc surface states, connecting two Weyl nodes of opposite chiralities. Therefore, equivalence between four (six) band model for the double (triple) Weyl fermions [see Eq.~(\ref{Eq:genWeyl_coupled_lattice})] and their two band models [see Eqs.~(\ref{eq:Weyl_2bandcriptic})-(\ref{eq:Ny_twoband})] can be established by comparing the number Fermi arcs for multi-Weyl systems from these two sets of tight-binding models. The results are shown in Figs.~\ref{Fig:Fermiarcs_2band} and \ref{Fig:Fermiarcs_multiband}.

To compute the Fermi arc surface states we impose periodic boundaries in the $y$ and $z$ directions, such that $k_y$ and $k_z$ can be treated as good quantum numbers. But, we implement open boundary in the $x$ direction, along which the linear dimensionality of the system is denoted by $L$~\cite{PhysRevB.96.201401,Nandy:2018fsi}. In such a mixed Bloch-Wannier representation the Fermi arcs are localized on the top and bottom surfaces, as shown in Figs.~\ref{Fig:Fermiarcs_2band} and \ref{Fig:Fermiarcs_multiband}. Specifically in Fig.~\ref{Fig:Fermiarcs_2band} we show the topologically protected Fermi arcs for multi-Weyl semimetals, constructed from their two band tight-binding models [see Eqs.~(\ref{eq:Weyl_2bandcriptic})-(\ref{eq:Ny_twoband})]. We find that a multi-Weyl semimetal, characterized by integer (anti-)monopole charge $n$, supports exactly $n$ copies of Fermi arc surface states. This observation establishes the bulk-boundary correspondence for this family of gapless topological semimetals. On the other hand, in Fig.~\ref{Fig:Fermiarcs_multiband} we show the Fermi arcs for double and triple Weyl semimetals, but constructed from the four and six band models [see Eqs.~(\ref{Eq:simpleweyl_TBexplicit}) and ~(\ref{Eq:genWeyl_coupled_lattice})], respectively. Once again we find that these two systems respectively host two and three copies of the Fermi arcs on the top and bottom surfaces. This outcome besides supporting the bulk-boundary correspondence, also anchors the topological equivalence between the multi and two band representations for the double and triple Weyl semimetals on a lattice. To appreciate some additional salient features of the arc states, next we consider their microscopic origin.

Any general Weyl semimetal hosting (anti-)monopole of charge $n$ can be constructed by stacking two-dimensional layers of quantum anomalous Hall insulator, occupying the $xy$-plane, in the momentum space along the $k_z$ direction within the range $-{\rm K}_0 \leq k_z \leq {\rm K}_0$, where ${\rm K}_0=\frac{\pi}{2}$ in our lattice construction. The first Chern number of each such anomalous Hall insulator is $n$ and it supports $n$ copies of one-dimensional chiral edge modes, with $n$ states at precise zero energy. The collection of such zero-energy states within the range $-{\rm K}_0 \leq k_z \leq {\rm K}_0$ constitutes $n$ copies of the Fermi arc surface states, shown in Figs.~\ref{Fig:Fermiarcs_2band} and \ref{Fig:Fermiarcs_multiband}. Also note that the localization length of each zero-energy mode is inversely proportional to bulk gap of the underlying two-dimensional anomalous Hall insulator for a given $k_z$. In our lattice models, such a gap is largest when $k_z=0$ and it vanishes at $k_z=\pm \frac{\pi}{2}$. Otherwise, this gap decreases smoothly as we approach $k_z=\pm \frac{\pi}{2}$ from the center of the surface Brillouin zone ($k_z=0$). Consequently, the surface localization of each copy of Fermi arcs decreases monotonically as we approach two Weyl points from the center of the arcs. Ultimately, at $k_z=\pm \frac{\pi}{2}$ the arcs are completely delocalized, and at these two points arcs from the top and bottom surfaces get connected via the bulk Weyl nodes. This feature can be seen from Figs.~\ref{Fig:Fermiarcs_2band} and \ref{Fig:Fermiarcs_multiband}. Next we proceed to derive the effective field theory of these systems.

\section{Effective field theory}~\label{Sec:QFT_MultiWeyl}

All global symmetries, present in a classical action, do not necessarily survive after quantization \cite{Bertlmann:1996xk}. Possibly the best known examples of this phenomenon are the ones related to the chiral anomalies. In particular, the axial anomaly is responsible for the celebrated decay of pion into two photons~\cite{Bell:1969ts, Adler}. Furthermore, it also leaves signatures on anomaly-induced transports that have attracted ample attention in recent time in the context of Dirac and Weyl semimetals in condensed matter systems, quark-gluon plasma in heavy-ion colliders and magnetized plasmas in cosmology, for example. In this section, we derive the effective field theory for multi-Weyl semimetals, unveil the anomaly structure therein and discuss its imprints on various transports.

After establishing the Hamiltonian description for multi-Weyl semimetals, we seek to formulate the corresponding Lagrangian formalism, which allows us to derive the effective field theory for these systems. Performing a Legendre transformation on the continuum Hamiltonian [see Eq.~(\ref{Eq:genWeyl_coupled})], we obtain the Lagrangian for left chiral fermions ($\psi_L$) 
 \begin{equation}
 \label{eq:DWL_Interaction}
 \mathcal{L}_{L} = i \psi^{\dagger}_{L} \tau^{\mu} \left[ \partial_{\mu} - i \Delta\left( \delta_{\mu}^x s_{x} +\delta_{\mu }^y s_{y}\right) \right] \psi_{L},
 \end{equation}\noindent
where $\tau^\mu=(1,\vec \tau)$. Throughout Einstein's summation convention over repeated indices is assumed. The above expression allows us to construct a generalized formalism for multi-Weyl semimetals with a non-Abelian $U(2)_L$ flavor symmetry, in presence of a non-Abelian background gauge field $A_{\mu}^as_a= \mathcal A_\mu^{0} s_0+\mathbb A_\mu^{i}s_i$ according to
 \begin{equation}
 \label{eq:DWL_Interaction_2}
 \mathcal{L}_{L} = i \psi^{\dagger}_{L} \tau^{\mu} \left[ \partial_{\mu} - i A^a_{\mu} s_a\right] \psi_{L},
 \end{equation}\noindent
where $s_a = (s_0,s_i)$ with $i=x,y,z$ are the generators of $U(1)_L\times SU(2)_L$. In particular, the static background field giving rise to nonlinear dispersion (in the $xy$ plane) in multi-Weyl systems can be written as 
\begin{equation}
A^a_{\mu}= \Delta (\delta_\mu^x \; \delta^{ax}+ \delta_\mu^y \; \delta^{a y}).
\end{equation}
From now on, we denote the field strength associated to $A_\mu$ by $F_{\mu\nu}$, the Abelian gauge field by ${\mathcal F}_{\mu \nu}$ and the $SU(2)$ gauge field by $G_{\mu\nu}$. Note that the non-Abelian field, giving rise to the multi-Weyl semimetals, picks a preferred direction and reduces the initial $SO(3,1)\times SU(2)_L$ symmetry group of $n$ decoupled copies of simple Weyl fermions to the diagonal $SO(1,1)\times U(1)_{3_L}$ symmetry.

Let us assume that we have a theory for left chiral fermions, transforming in some representation $\mathcal{R(G)}$ of the Lie group $\mathcal{G}$. The corresponding generators $s_a$ of the Lie algebra satisfy  
\begin{equation}
\left[s_a,s_b\right] = if^{abc}s_c,
\end{equation}
where $f^{abc}$ is the structure factor of the Lie group ${\mathcal G}$. For theories with such a {\em flavor} symmetry the associated anomalous currents ($J_a^{\mu}$) satisfy the following Ward identities in its covariant form
\begin{eqnarray}~\label{eq:anomalies}
 \mathcal{D}_{\mu} J_a^{\mu}  &=& \frac{d_{abc}}{32 \pi^2} \; \epsilon^{\mu \nu \rho \lambda} \; F^{b}_{\mu \nu} \; F^c_{\rho \lambda} \nonumber \\
  & +& \frac{b_a}{768 \pi^2} \; \epsilon^{\mu \nu \rho \lambda} \; R^{\alpha}\,_{\beta \mu \nu} \; R^{\beta}\,_{\alpha \rho \lambda}, \\
\nabla_\mu T^{\mu}\,_\nu &=& F^a_{\nu\mu} J_a^\mu + \frac{b_a}{384\pi^2} \; \mathcal{D}^\mu\left(\epsilon^{\sigma \kappa \rho \lambda} \; F^a_{\kappa\sigma} \; R_{\nu\mu \rho \lambda}\right), \nonumber \\
\end{eqnarray}
where $\mathcal{D}_{\mu}$ is the covariant derivative containing the gauge and metric connections, $\nabla_\mu$ is the curved space covariant derivative, $R_{\nu\mu \rho \lambda}$ is the Rieman curvature tensor, $T^{\mu\nu}$ is the stress-energy tensor, and the anomalous coefficients are $d_{abc}= \frac{1}{2} \mathrm{Tr} \left[ \left\lbrace s_a,s_b\right\rbrace  s_c\right] $ and $b_{a}= \mathrm{Tr} \left[ s_a\right]$  \cite{Bertlmann:1996xk,Landsteiner:2012kd}\footnote{For right chiral fermions $d_{abc}= -\frac{1}{2} \mathrm{Tr} \left[ \left\lbrace s_a,s_b\right\rbrace  s_c\right] $ and $b_{a}= -\mathrm{Tr} \left[ s_a\right]$.}. The currents $J_a^\mu$ cannot be obtained by varying an action with respect to the background fields $A^a_\mu$, and normally it is called {\em covariant currents}. Nonetheless, there exist the {\em consistent currents} $\tilde J_a^\mu$ related with $J^\mu_a$ by the addition of a Chern-Simons polynomial. The consistent currents can be defined as a functional derivative of the action with respect to the background field according to $\tilde J_a^\mu=\delta W/\delta A_\mu^a$, where $J_a^\mu = \tilde J_a^\mu + K_a^\mu$ and 
\begin{equation}~\label{eq:BZPoly}
K_a^\mu = -\frac{1}{48\pi^2}\epsilon^{\mu \nu \rho \lambda} \; \mathrm{Tr}\left[s_a\left(\{A_\nu, F_{\rho\lambda}\} -A_\nu A_\rho A_\lambda \right)\right].
\end{equation}
 Thus, the consistent Ward identity reads
\begin{eqnarray}~\label{eq:consWI}
{\mathcal D}_\mu \tilde J^\mu_a &=& \frac{1}{24\pi^2}\epsilon^{\mu\nu\rho\sigma} \; \mathrm{Tr} \big[ s_a\partial_\mu \big( A_\nu\partial_\rho A_\sigma \\
&+& \frac{1}{2}A_\nu A_\rho A_\sigma \big) \big]
+ \frac{b_a}{768 \pi^2} \; \epsilon^{\mu \nu \rho \lambda} R^{\alpha}\,_{\beta \mu \nu}R^{\beta}\,_{\alpha \rho \lambda}. \nonumber
\end{eqnarray}

To illustrate the applicability of the general theoretical framework discussed so far, we now focus on a theory with one copy of left and right handed fermions, coupled to Abelian gauge fields. The Ward indentities for the consistent current $\tilde J_e^\mu = \tilde J_L^\mu+\tilde J_R^\mu$ that couples to the gauge field $\mathcal A_\mu=\frac{1}{2}(\mathcal A^L_\mu+\mathcal A^R_\mu)$ and the axial current, defined as $\tilde J_5^\mu = \tilde J_L^\mu-\tilde J_R^\mu$, respectively read  
\begin{align}\label{eq:wIA1}
	\nabla_{\mu}\tilde J_e^{\mu} {}&= \frac{1}{32 \pi^2} \epsilon^{\mu \nu \rho \lambda} \mathcal F_{\mu \nu} \mathcal F^{5}_{\rho \lambda}, \\
	\nabla_{\mu}\tilde J_{5}^{\mu} {}&= \frac{1}{48 \pi^2} \epsilon^{\mu \nu \rho \lambda} \left( \mathcal F_{\mu \nu} \mathcal F_{\rho \lambda}
	+ \mathcal F^{5}_{\mu \nu} \mathcal F^{5}_{\rho \lambda}\right)\nonumber \\ 
	&+\frac{1}{768 \pi^2}\epsilon^{\mu \nu \rho \lambda} R^{\alpha}\,_{\beta \mu \nu}R^{\beta}\,_{\alpha \rho \lambda}\,.
\end{align}  
The conservation of electric charge requires that the combination $U(1)_{L} + U(1)_{R} \equiv U(1)_e$ should be conserved. However, in Eq. \eqref{eq:wIA1} the vector current $\tilde J_e^\mu$ is not conserved\footnote{Notice that Eq. \eqref{eq:wIA1} is an operator equation. Therefore, even though at the fundamental level axial gauge fields do not exist, the three point function $\langle\partial_\mu\tilde J_e^{\mu}\tilde J_e^{\nu}\tilde J_5^{\rho}\rangle\neq 0$ spoils the consistency of the theory.}. Nonetheless, this issue can be resolved by noting that the theory is not gauge invariant. Hence, one can add a counter-term, known as the `Bardeen counter-term' 
\begin{equation}
W_{\rm BCT} = -\frac{1}{12\pi^2}\int d^4x \epsilon^{\mu\nu\rho\lambda}A^e_\mu A^5_\nu F_{\rho\lambda}\,,
\end{equation}
to the original action, which reestablishes the $U(1)_e$ gauge invariance. After introducing this local polynomial, the Ward identities for the newly defined consistent currents 
\begin{equation}
j_{e} = \tilde J_{e} + \frac{\delta W_{\rm BCT}}{\delta A_\mu^{e}}, \:\:\: 
j_{5} = \tilde J_{5} + \frac{\delta W_{\rm BCT}}{\delta A_\mu^{5}}, 
\end{equation}
respectively read as
\begin{align}~\label{eq:wIA2}
	\nabla_{\mu} \; j_e^{\mu} {}&= 0 \\
	\nabla_{\mu} \; j_{5}^{\mu} {}&= \frac{1}{16 \pi^2} \epsilon^{\mu \nu \rho \lambda} \left( \mathcal F_{\mu \nu} \mathcal F_{\rho \lambda} + \frac{1}{3}\mathcal F^{5}_{\mu \nu} \mathcal F^{5}_{\rho \lambda}\right)\\
	&+\frac{1}{768 \pi^2}\epsilon^{\mu \nu \rho \lambda} R^{\alpha}\,_{\beta \mu \nu}R^{\beta}\,_{\alpha \rho \lambda}\,.  \nonumber
\end{align}  
Only after such a redefinition, the current $j^\mu_e$ associated with the $U(1)_e$ electric charge is conserved, while the axial current $j^\mu_5$ remains anomalous. The above construction has a natural generalization to the theories with non-Abelian anomalies, which we discuss next.

\subsection{Theory with $U(1) \times SU(2)$ flavor symmetries}

Now we consider multi-Weyl semimetals, in which the left- and right-handed fermions transform under a $SU(2)_{L/R}$ representation. In this case, the (covariant) anomalous Ward identities [see Eq.~\eqref{eq:anomalies}] read as follows
\begin{align}
\nabla_{\mu} J_e^{\mu} = {}& \frac{n}{8 \pi^2}  \epsilon^{\mu \nu \rho \lambda} \mathcal F_{\mu \nu}\mathcal F^{5}_{\rho \lambda} + \frac{c(n)}{8 \pi^2}  \epsilon^{\mu \nu \rho \lambda}  {G_{\mu \nu}} ^i {G^{5}_{\rho \lambda} }^i \, , \\
\nabla_{\mu} J_{5}^{\mu}= {}& \frac{n}{16 \pi^2}  \epsilon^{\mu \nu \rho \lambda} \left(\mathcal F_{\mu \nu} \mathcal F_{\rho \lambda}+\mathcal F^{5}_{\mu \nu} \mathcal F^{5}_{\rho \lambda} \right) \nonumber \\
{}& +\frac{c(n)}{16 \pi^2}  \epsilon^{\mu \nu \rho \lambda} \left({G_{\mu \nu}} ^i {G_{\rho \lambda} }^i+{G^{5}_{\mu \nu}} ^i {G^{5}_{\rho \lambda} }^i \right)  \nonumber\\
{}&+\frac{n}{384 \pi^2}\epsilon^{\mu \nu \rho \lambda} R^{\alpha}_{\beta \mu \nu}R^{\beta}_{\alpha \rho \lambda} \,,\\
\mathcal D_{\mu} J_i^{\mu} = {}& \frac{c(n)}{8 \pi^2}  \epsilon^{\mu \nu \rho \lambda}\left( \mathcal F_{\mu \nu}G^{5}_{\rho \lambda}\,^i + \mathcal F^5_{\mu \nu}G_{\rho \lambda}\,^i\right) \, , \\
\mathcal D_{\mu} J_{i,5}^{\mu}= {}&  \frac{c(n)}{8 \pi^2}  \epsilon^{\mu \nu \rho \lambda}\left( \mathcal F_{\mu \nu}G_{\rho \lambda}\,^i + \mathcal F^5_{\mu \nu}G^5_{\rho \lambda}\,^i\right) \,.
\end{align} \noindent
where $c(n)$ is defined via the relation $\mathrm{Tr}(s_i s_j)= c(n)\delta_{ij}$ for $n>1$. For our choices of the generators, $c(2)=1/2$ and $c(3)=2$. As discussed in the previous section, the covariant current cannot be obtained by differentiating any functional of the gauge fields. Therefore, they do not couple to the gauge fields. However, Bardeen computed the proper counter-term to construct conserved vector consistent currents~\cite{PhysRev.184.1848}\footnote{We impose the conservation of the vector non-abelian current because the continuum version of the lattice $C_4$ symmetry corresponds to the $U(1)_{3_e}=U(1)_{3_L}+ U(1)_{3_R}$, discussed in the previous section.}. Combining the Bardeen counter-terms with the Bardeen-Zumino polynomial [see Eq.~\eqref{eq:BZPoly}] \cite{Manes:2018llx,Manes:2018mth}, we can write the Chern-Simons current, relating the covariant and consistent currents according to
\begin{equation}
J^\mu_a = j^\mu_a + P^\mu_a,
\end{equation}
where
\begin{eqnarray}\label{eq:ChernSPoly}
P_a^\mu &=& \frac{1}{8\pi^2}\epsilon^{\mu\nu\rho\lambda} \; {\rm Tr} \; \big[ s_a \big( A^5_\nu F_{\rho\lambda}+F_{\rho\lambda} A^5_\nu \nonumber \\
&+& \frac{8}{3}iA^5_\nu A^5_\rho A^5_\lambda \big) \big]\,,\\
P_{a,5}^\mu &=& \frac{1}{24\pi^2}\epsilon^{\mu\nu\rho\lambda} \; {\rm Tr} \; \left[s_a\left(A^5_\nu F^5_{\rho\lambda}+F^5_{\rho\lambda} A^5_\nu\right)\right]\,.
\end{eqnarray}
Having understood the anomaly structure of the effective field theory, we now extract the anomaly-induced transport coefficients for multi-Weyl systems.

\begin{figure*}[t!]
\subfigure[]{
\includegraphics[width=0.48\linewidth]{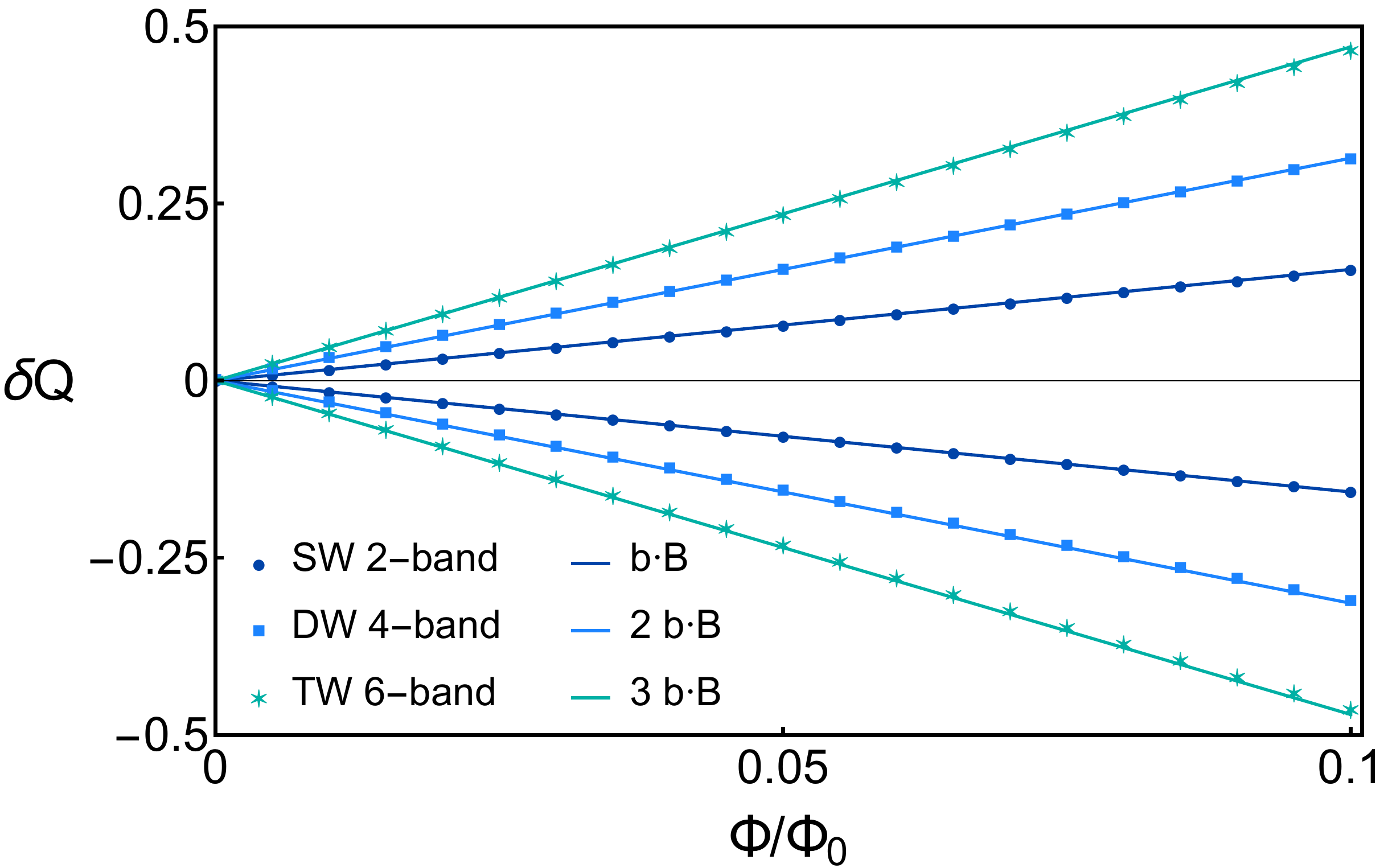}~\label{Fig:abeliananomaly_multiband}
}
\subfigure[]{
\includegraphics[width=0.48\linewidth]{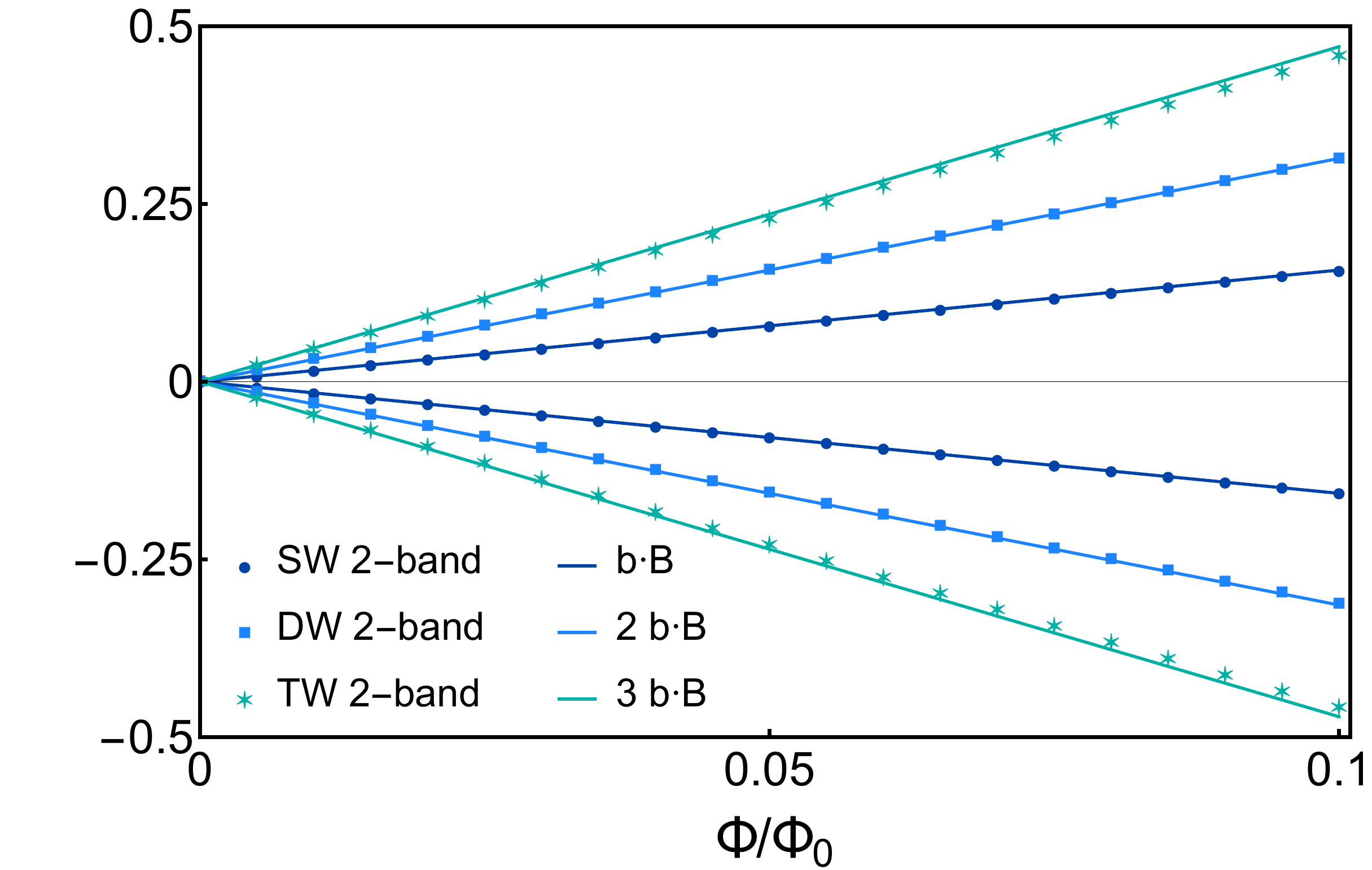}~\label{Fig:abeliananomaly_twoband}
}
\caption{ The charge density ($\delta Q$), measured in units of $e/\pi$, in the vicinity of the flux tubes placed at $x=L/4$ (yielding $\delta Q>0$) and $x=-L/4$ (yielding $\delta Q<0$) [see Eq.~(\ref{Eq:Bprofile})] for simple Weyl (SW), double Weyl (DW) and triple Weyl (TW) semimetals. Results for simple Weyl semimetal are always obtained from the two band model. By contrast, for the double (triple) Weyl semimetals, we compute $\delta Q$ from (a) four (six) and (b) two band models. Here, $\Phi$ is the magnetic flux, measured in units of the flux quanta $\Phi_0=2 \pi /e$ (in natural units $\hbar=c=1$). For numerical analysis, we always take $t=t_z=t_0=1$ and $L_z=200$, where $L_z$ is the number of points in the momentum space along the $k_z$ direction. Specifically for (a) we set $\Delta=0.2$ and  $L=24$, while for (b) $L=48$, where $L$ is the linear dimension of the system in the $x$ and $y$ directions. The dots correspond to numerically computed values of $\delta Q$, whereas the straight lines follow the relation $\delta Q= n  \frac{e}{2\pi} ({\bf b} \cdot {\bf B}) \equiv n \; b \; (\Phi/\Phi_0)$, where $n$ is the monopole charge of the Weyl nodes, located at ${\bf b}=(0,0,\pm b)$, and $b=\pi/2$. This analysis establishes excellent agreement between the field theoretic predictions and the scaling of the corresponding observable computed from the lattice models.   
}~\label{Fig:AbelianAnomaly_Summary}
\end{figure*}

\subsection{Anomaly induced transport}\label{sec:4b}

Chiral fermions exhibit non-dissipative transports at finite temperature ($T$) and density ($\mu$), which are intimately related to the chiral anomalies discussed in the previous sections. In particular, the covariant currents within the linear response approximation were computed in Refs.~\cite{Son:2009tf,Kharzeev:2009pj,Neiman:2010zi,Landsteiner:2011cp,PhysRevD.20.1807,PhysRevD.22.3080,PhysRevD.99.056003,Loganayagam:2012pz,2013JHEP...02..088J} and read~\footnote{These expressions assume linear response, therefore in the definition of the non-abelian magnetic fields only the linear terms in the gauge fields are considered.}
\begin{eqnarray}
J^\mu_a &=& \sigma^B_{ab} \; B^\mu_b + \sigma^V_a \; \omega^\mu, \\
T^{\mu\nu} &=& \sigma_a^{\epsilon,B} \; u^{(\mu} \; B^{\nu)}_a+ \sigma^{\epsilon,V} \; u^{(\mu} \; \omega^{\nu)}\,,
\end{eqnarray}
where $A^{(\mu} C^{\nu)} =(A^\mu C^\nu + A^\nu C^\mu)/2$. The magnetic and vorticity fields respectively are defined as
\begin{equation}
B^{\nu}_a =\epsilon^{\mu\nu\rho\lambda}u_\nu \nabla_\rho A^a_{\lambda}\quad,\quad \omega^\mu = \epsilon^{\mu\nu\rho\lambda}u_\nu\nabla_\rho u_\lambda,
\end{equation} 
 and $u^\mu$ is a unit norm timelike vector. The nondissipative currents give rise to (a) chiral magnetic conducitivities ($\sigma^B_{ab}$ and $\sigma_a^{\epsilon,B}$) and (b) chiral vortical conductivities ($\sigma^V_a$ and $\sigma^{\epsilon,V}$). In the absence of dynamical gauge fields, these quantities are universal and solely determined by the anomaly. And they are given by
\begin{eqnarray}
\sigma^B_{ab} &=& \frac{1}{4\pi^2} \; d_{abc} \; \mu^c, \\
\sigma^V_a &=&\sigma^{\epsilon,B}_a = \frac{1}{8\pi^2} \; d_{abc} \; \mu^b \; \mu^c + \frac{T^2}{24} \; b_a, \\
\sigma^{\epsilon,V} &=& \frac{1}{12\pi^2} \; d_{abc} \; \mu^a\mu^b\mu^c+ \frac{T^2}{12} \; b_a \; \mu^a,
\end{eqnarray}
where $\mu^a=(\mu, \mu^i)$, with $\mu$ and $ \mu^i$ denoting the regular and flavor chemical potentials, respectively.

For a theory with $U(1)_e \times U(1)_5$ symmetry (describing a simple Weyl semimetal with $n=1$), the vector and axial covariant currents are respectively given by 
\begin{eqnarray}
{\bf J}_e &=& \frac{\mu_5}{2\pi^2} \; {\bf B}  +\frac{\mu}{2\pi^2} \; {\bf B}_5 + \frac{\mu\;\mu_5}{2\pi^2} \; {\boldsymbol \omega}, \\
{\bf J}_5 &=& \frac{\mu}{2\pi^2} \; {\bf B}  +\frac{\mu_5}{2\pi^2} \; {\bf B}_5+ \left( \frac{\mu^2+\mu_5^2}{4\pi^2} + \frac{T^2}{12}\right) \; {\boldsymbol \omega}.
\end{eqnarray}
On the other hand, the covariant current arising from the energy-momentum tensor, $J^{i}_{\epsilon}=T^{i0}$, is
\begin{eqnarray}
{\bf  J}_\epsilon &=& \frac{\mu \; \mu_5}{2\pi^2} \; {\bf B}+ \left(\frac{\mu^2+\mu_5^2}{4\pi^2} + \frac{T^2}{12}\right) \; {\bf B}_5 \nonumber \\
&+& \left(\frac{\mu_5}{6\pi^2}\left(3\mu^2+\mu_5^2\right)+ \frac{\mu_5T^2}{6}\right) \; {\boldsymbol \omega}.
\end{eqnarray}
In the above expressions, we introduce the following quantities $\mu=\mu_{_R}+\mu_{_L}$, $\mu_5=\mu_{_L}-\mu_{_R}$, $\mathcal A_\mu^L=\mathcal A_\mu+\mathcal A_\mu^5$ and $\mathcal A_\mu^R=\mathcal A_\mu-\mathcal A_\mu^5$. However, as discussed in  Sec. \ref{Sec:QFT_MultiWeyl} the covariant electric current does not couple to the electromagnetic field. The proper conserved current that couples to the photon is the consistent current, obtained after including the contribution from the Bardeen-Zumino polynomial and Bardeen couternterm [see Eq.~\eqref{eq:ChernSPoly}], leading to the following expressions
\begin{eqnarray}
\rho_e &=& \frac{1}{2\pi^2} \; {\bf b} \cdot {\bf B},  \\
{\bf j}_e &=& \frac{\mu}{2\pi^2} \; {\bf B}_5 + \frac{\mu \; \mu_5}{2\pi^2} \; {\boldsymbol \omega} +  \frac{1}{2\pi^2} \; {\bf E} \times {\bf b}, \\
\rho_5 &=& \frac{1}{6\pi^2} \; {\bf b} \cdot {\bf B}_5, \\
{\bf  j}_5 &=& \frac{\mu}{2\pi^2} \; {\bf B} +\frac{\mu_5}{3\pi^2} \; {\bf B}_5 + \left( \frac{\mu^2+\mu_5^2}{4\pi^2} + \frac{T^2}{12}\right)\; {\boldsymbol \omega} \nonumber \\
&+& \frac{1}{6\pi^2} \; {\bf E}_5 \times {\bf b}\,.\label{eq:abj}
\end{eqnarray}
While arriving at the final expression, we have used $\mathcal A^5_\mu=(\mu_5,{\bf b})$, in order for the model to describe a time reversal symmetry breaking Weyl semimetal. In this construction, the separation of two Weyl nodes is $2 |{\bf b}|$. Note that the covariant current (${\bf J}_e$) has a contribution $\sim \mu_5 {\bf B}$, which captures static chiral magnetic effect. However, the Chern-Simons current contains a contribution $-A^5_0 {\bf B} \equiv -\mu_5 {\bf B}$, which exactly cancels such contribution in the conserved current. Therefore, Weyl systems do not exhibit any static chiral magnetic effect in the vector current (${\bf j}_e$)~\cite{Landsteiner:2016led}.

After establishing the current operators for the Abelian field theory with $U(1)_e \times U(1)_5$ symmetry, we now construct both the Abelian and non-Abelian currents from the field theoretic description of the multi-Weyl systems, possessing a $U(1)_e \times U(1)_5 \times SU(2)_e \times SU(2)_5$ symmetry. At this stage, we introduce the notion of the isospin chemical potentials $\mu_{3}=\mu^3_{_L}+\mu^3_{_R}$ and $\mu_{3_5}=\mu^3_{_L}-\mu^3_{_R}$, respectively and the corresponding vector and axial gauge fields $\mathbb A^{3,L}_\mu=\mathbb A^3_\mu+\mathbb A_\mu^{3_5}$ and $\mathbb A^{3,R}_\mu=\mathbb A^3_\mu-\mathbb A_\mu^{3_5}$. The covariant Abelian currents now read as
\begin{eqnarray}
{\bf J}_e &=& \frac{n}{2\pi^2} \left( \mu_5 {\bf B} + \mu {\bf B}_5 \right)  + \frac{c(n)}{2\pi^2} \left(\mu_{3_5} {\bf B}_3 + \mu_{3} {\bf B}_{3_5} \right), \nonumber \\
{\bf J}_5 &=& \frac{n}{2\pi^2} \left( \mu {\bf B} +  \mu_5 {\bf B}_5  \right) + \frac{c(n)}{2\pi^2} \left(\mu_{3} {\bf B}_3 + \mu_{3_5} {\bf B}_{3_5} \right). 
\end{eqnarray}
For the sake of simplicity, we here ignore the contribution from the vortical conductivities and the energy current, which we show in the appendix~\ref{Append:vorticalnonAb}. To arrive at the consistent currents, we need to evaluate the Chern-Simons polynomial [see Eq.~\eqref{eq:ChernSPoly}] and add it to the above covariant currents. In particular, for the Abelian charge densities and currents we obtain 
\allowdisplaybreaks[4]
\begin{eqnarray}
\rho_e &=& \frac{n}{2\pi^2} \; {\bf b} \cdot {\bf B}, \\
{\bf j}_e &=& \frac{n}{2\pi^2} \mu {\bf B}_5 + \frac{c(n)}{2\pi^2}\mu_{3} {\bf B}_{3_5} +  \frac{n}{2\pi^2} {\bf E} \times {\bf b},\\
\rho_5 &=& \frac{n}{6\pi^2} \; {\bf b} \cdot {\bf B}_5\\
{\bf j}_5 &=& \frac{n}{2\pi^2}\mu {\bf B} +\frac{n}{3\pi^2}\mu_5 {\bf B}_5 + \frac{c(n)}{2\pi^2}\mu_{3} {\bf B}_3+ \frac{c(n)}{3\pi^2}\mu_{3_5} {\bf B}_{3_5} \nonumber \\
&+&  \frac{n}{6\pi^2} {\bf E}_5 \times {\bf b}\,.
\end{eqnarray}
On the other hand, the non-Abelian densities and currents take the following forms
\allowdisplaybreaks[4]
\begin{eqnarray}
\rho_3 &=& \frac{c(n)}{2\pi^2} \; {\bf b} \cdot {\bf B}_3,~\label{eq:abelcurrent1} \\
{\bf j}_3 &=&   \frac{c(n)}{2\pi^2} \; \mu {\bf B}_{3_5} +\frac{c(n)}{2\pi^2} \; \mu_3 {\bf B}_5+ \frac{c(n)}{2\pi^2} \; {\bf E}_3 \times {\bf b}, \\
\rho_{3_5} &=& \frac{c(n)}{2\pi^2}\; {\bf b} \cdot {\bf B}_{3_5}+ \frac{c(n)}{6\pi^2} \; {\bf b} \cdot {\bf B}_{3_5}\\
{\bf j}_{3_5} &=& \frac{c(n)}{2\pi^2} \; \mu_3 {\bf B} +  \frac{c(n)}{3\pi^2} \; \mu_{3_5} {\bf B}_{5}+ \frac{c(n)}{2\pi^2} \; \mu {\bf B}_3  \nonumber \\
 &+&\frac{c(n)}{3\pi^2}\; \mu_{5} {\bf B}_{3_5}+   \frac{c(n)}{6\pi^2} \; {\bf E}_{3_5} \times {\bf b}.~\label{eq:nabelcurrent1}
\end{eqnarray}
In order to relate these currents with the multi-Weyl semimetal we need to take into account the presence of the non-Abelian background field $\mathbb A=\Delta\left(0, s_x, s_y,0\right)$. The presence of such background field introduces a non-Abelian magnetic field ${\bf B}_3=(0,0,\Delta^2)$. 

However, two issues associated to such a background field that we should address. The first one is associated to the fact that the transport coefficients shown before were computed using the linear response theory and the actual magnetic field correspond to a non-linear contribution, and secondly the parameter $\Delta$ also breaks the $SU(2)$ symmetry spoiling the anomaly protection of the current, shown in Eqs.~(\ref{eq:abelcurrent1})-(\ref{eq:nabelcurrent1}). We deal this this subtlety in Sec.~\ref{sec:HOLO}. Prior to that, we proceed to anchor some of the predictions from the effective field theory for multi-Weyl semimetals from their lattice realizations.


\section{Anomalous Responses from Lattice models}~\label{Sec:QFT_MultiWeyl_Lattice}

In the previous sections, we established the low-energy models (both in the continuum and from the tight-binding models on a cubic lattice) and the effective field theoretic description for multi-Weyl semimetals. We now test the predictions from the effective field theory (see Sec.~\ref{Sec:QFT_MultiWeyl}) by computing some specific observables or expectation values of some operators from the lattice regularized models, introduced in Sec.~\ref{subsec:onlylatticemodels}. We first focus on the Abelian sector, for which the anomalous Hall effect yields the following relations between the charge ($\rho_e$) and current ($\bf j_e$) densities 
\begin{equation}~\label{eq:Hallanomaly_abelian}
\rho_e= n \; \frac{e^2}{2 \pi^2} \; \left( \mathbf{b} \cdot \mathbf{B} \right), 
\qquad \mathbf{j}_e= n \; \frac{e^2}{2 \pi^2} \; \left( \mathbf{b} \times \mathbf{E} \right),
\end{equation}
where $n$ is the monopole charge, $e$ is the Abelian electric charge, ${\bf B} ({\bf E})$ is the external magnetic (electric) field, and the Weyl points are located at ${\bf b}=(0,0,\pm b)$. In our lattice models, $b=\pi/(2 a)$, where $a$ is the lattice spacing, set to be unity for convenience.

Since the charge and current densities respectively arise from the temporal and spatial components of the same Chern-Simons current, we only compute $\rho_e$ from the lattice model. To this end, we consider a cubic lattice, with $L$ sites in the $x$ and $y$ directions and preserve translational invariance along the $z$ direction (leaving $k_z$ as a good quantum number). We impose periodic boundaries in the $x$ and $y$ directions. The effect of the magnetic field can be incorporated via the Peierls substitution: hopping terms between lattice sites at $\mathbf{r}_i$ and $\mathbf{r}_f$ acquire the phase $\exp\left[{\frac{2\pi i}{\Phi_0} \int_{\mathbf{r}_i}^{\mathbf{r}_f} \mathbf{A}\cdot d\mathbf{r}} \right]$, where $\mathbf{A}$ is the Abelian vector potential and $\Phi_0=h c/e$ is the flux quanta. In what follows, the magnetic field assumes the following profile~\cite{PhysRevLett.111.027201}
\begin{equation}~\label{Eq:Bprofile}
\mathbf{B}= \Phi \; \left[\delta \left( x-\frac{L}{4} \right)-\delta \left( x + \frac{L}{4} \right) \right] \delta(y) \; \hat{z},
\end{equation}
where $\Phi$ is the flux produced by the external magnetic field and $\delta$ is the Dirac delta function. For convenience, we work in the Landau gauge, given by 
\begin{equation}
{\bf A}=\Phi \; \left[\Theta \left( x-\frac{L}{4} \right)-\Theta \left( x + \frac{L}{4} \right) \right] \delta(y) \; \hat{y},
\end{equation}
where $\Theta$ is the heaviside step function.

Upon numerically diagonalizing the tight-binding model in the presence of such a singular magnetic field, we compute the charge accumulation at a given point $(x,y)$ from the following expression 
\begin{equation}
\rho_e(x,y)= e \sum_{E_i<0} \sum_{\alpha} \sum_{k_z} \left| \Psi_{\alpha}(x,y,k_z, E_i) \right|^2,
\end{equation}
where $\alpha$ is the pseudospin and flavor multi-index, $\Psi_\alpha(x,y,k_z, E_i)$ is the eigenstate with energy $E_i$ and the summation is performed over the ground state configuration, hence $E_i <0$. In order to perform the summation over $k_z$ we discretize the interval $k_z \in (-\pi, \pi)$ into $L_z$ points. The accumulated charge density around the flux tube located at $x=L/4$ is given by 
\begin{equation}
\delta Q= \sum^{L/2}_{x=1} \: \sum^{L}_{y=1} \:\: \frac{\rho_e(x,y)}{L_z}. 
\end{equation}       
The scaling of $\delta Q$ with $\Phi$ for multi-Weyl semimetals are shown in Fig.~\ref{Fig:AbelianAnomaly_Summary}.

\begin{figure}[t!]
\includegraphics[width=0.95\linewidth]{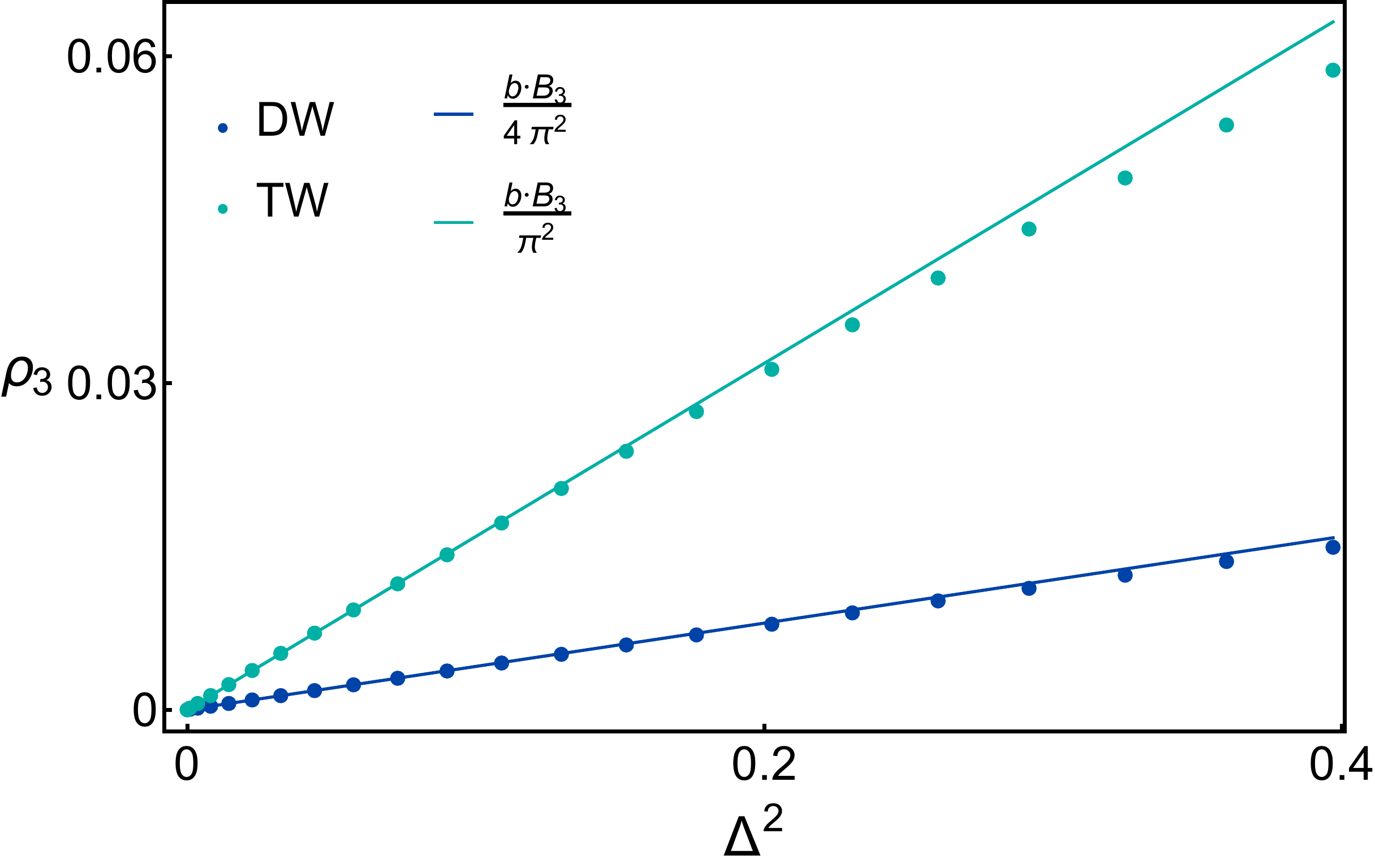}
\caption{The non-Abelian or isospin density ($\rho_{3}$) as a function of the non-Abelian magnetic field ($\Delta^2$), computed numerically from the four and six band models for the double Weyl (DW) and triple Weyl (TW) semimetals, respectively. For numerical analysis we set $t=t_z=t_0=1$ and discretize each momentum direction into 41 points in the Brillouin zone. In our construction, $|{\bf B}_3|=\Delta^2$ and $b=\pi/2$.
}~\label{Fig:nonAbelian_summary}
\end{figure}

We now compare the outcomes with the field theoretic predictions. In the natural units ($\hbar=c=1$), the flux quantum is $\Phi_0=2\pi/e$, and when we measure the accumulated charge density in units of $e/\pi$, then the anomaly equation for $\rho_e$ or $\delta Q$ from Eq.~(\ref{eq:Hallanomaly_abelian}) becomes 
\begin{equation}
\delta Q= n \; b \: \left( \frac{\Phi}{\Phi_0} \right). 
\end{equation} 
Note that in numerical analysis $\rho$ or $\delta Q$ is measured with respect to its expectation value in the absence of the external magnetic field. For simple Weyl semimetals we always compute this quantity from the two band model. However, for double (triple) Weyl semimetals, we compute this quantity from the four (six) [see Fig.~\ref{Fig:abeliananomaly_multiband}] and two band [see Fig.~\ref{Fig:abeliananomaly_twoband}] models. From Fig.~\ref{Fig:AbelianAnomaly_Summary}, we find that the slope of the straight lines in the $(\Phi/\Phi_0, \delta Q)$ plane is given by $(n \pi/2)$ for weak enough magnetic fields ($B a^2 \ll 1$), irrespective of the microscopic details. These results establish an excellent agreement with the field theoretic predictions.

Next we turn our focus on the anomalous Hall effect for the non-Abelian density ($\rho_3$) and current (${\bf J}_3$), given by 
\begin{equation}
\rho_3=  \frac{c(n)}{2 \pi^2} \; \left( \mathbf{b} \cdot \mathbf{B}_3 \right), \qquad \mathbf{j}_3= \frac{c(n)}{2 \pi^2} \;  \left(\mathbf{E}_3 \times \mathbf{b} \right), 
\end{equation}
where ${\bf B}_3$ (${\bf E}_3$) is the non-Abelian magnetic (electric) field, pointing in the $z$-direction in the flavor/isospin space, and $c(n)=1/2 \; (2)$ for $n=2\;(3)$. Recall that the above set of non-Abelian anomaly equations are only germane for multi-Weyl semimetals with $n>1$. To test the validity of the field theoretic predictions, we compute the non-Abelian or isospin density $\rho_3$ from the lattice models for double and triple Weyl fermions. Note that presence of the non-Abelian field is transparent only in the four (six) band models for double (triple) Weyl semimetals, and in particular ${\bf B}_3=(0,0,\Delta^2)$. Thus, we compute $\rho_3$ from these models only. Since ${\bf B}_3$ is an intrinsic homogenous field, the computation is performed in the momentum space representation of the corresponding lattice models. In particular, $\rho_3$ is computed from the following expression 
\begin{equation}
\rho_3= \frac{1}{L_x L_y L_z} \sum_{E_i <0} \sum_{{\bf k}} \langle \Psi \left(E_i,{\bf k} \right) \left| \tau_0 \otimes s^n_3  \right| \Psi \left(E_i,\mathbf{k} \right) \rangle,
\end{equation}
where $L_j$ is the number of sites in the $k_j$ direction in the momentum space, where $j=x,y,z$. The results are displayed in Fig.~\ref{Fig:nonAbelian_summary}.

Numerically evaluated $\rho_3$ is compared with $\Delta^2$ (strength of the non-Abelian magnetic field ${\bf B}_3$), and we find excellent linear dependence of $\rho_3$ on $\Delta^2$ for $n=2$ and $3$ for small $\Delta^2$, such that $\Delta a \ll 1$. The slopes of the linear fits are respectively given by $b/(4\pi^2) \equiv 1/(8 \pi^3)$ and $b/(\pi^2) \equiv 1/(2 \pi^3)$ for $n=2$ and $3$, since in our lattice construction $b=\pi/2$. This observation establishes an excellent agreement between the field theoretic predictions on the non-Abelian anomaly for multi-Weyl semimetals with the numerical findings from the lattice models.

In the context of non-Abelian anomaly a comment is due at this stage. We note that the isospin density operator $\tau_0 \otimes s^n_3$ defined in the four (six) band models for double (triple) Weyl semimetal, reduces to $\sigma_3$ in their low-energy sector (emergent two band description). In principle, we can compute the expectation value for this operator $\sigma_3$ from the two band models for the double and triple Weyl semimetals. However, note that the operator $\sigma_3$ can be obtained by projecting multiple operators defined in four or six band model. Hence, \emph{inverse of the projection operation is not unique}. Consequently, even though the expectation value of $\sigma_3$, namely $\langle \sigma_3 \rangle$, is finite, we find that in general it is much larger than the pure non-Abelian anomaly contribution. Presently, there is no known procedure to isolate the contribution in $\langle \sigma_3 \rangle$ arising purely from the non-Abelian anomaly. This is the reason we do not display the results on the scaling of $\langle \sigma_3 \rangle$ with $|{\bf B}_3|=\Delta^2$.


\section{Anomalous transport from holography}~\label{sec:HOLO}

\begin{figure}[t!]
	\centering
	\includegraphics[width=0.4\textwidth]{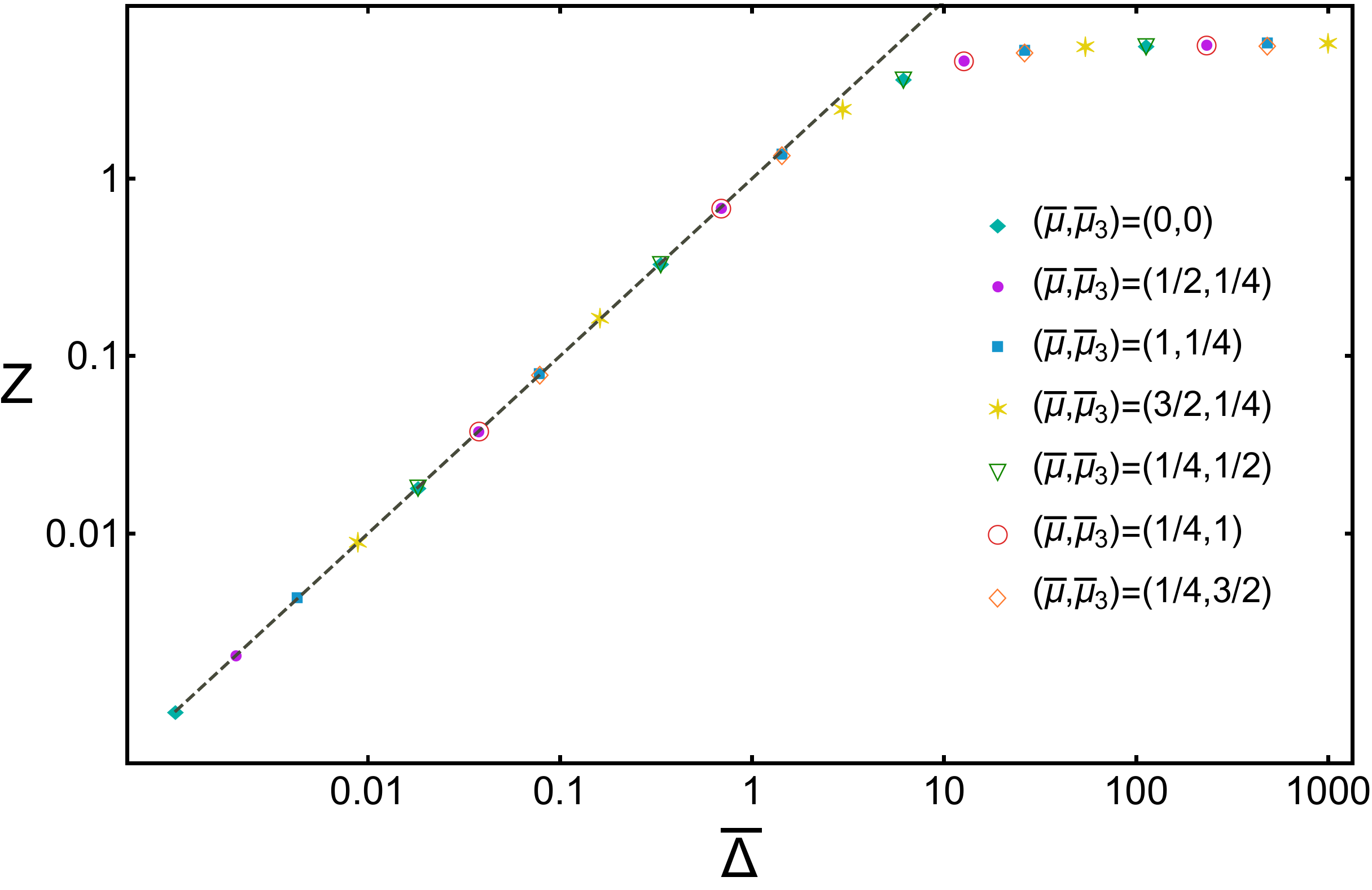}
	\caption{Renormalization of the background non-abelian  gauge field for different chemical potentials (see inset) as a function of $\bar \Delta$. The IR values correspond to $\bar \Delta\to\infty$. The dashed line corresponds to a linear fit $Z = \bar \Delta$.}
	\label{fig:bIR}
\end{figure}

As discussed in Sec.~\ref{Sec:QFT_MultiWeyl}, the presence of the background non-Abelian gauge field breaks the original global symmetry group. For the sake of simplicity, we consider only left-handed matter fields. In this case, the starting symmetry group is $SO(3,1)\times SU(2)_L\times U(1)_L$, which gets broken down to $SO(1,1)\times U(1)_{3_L}\times U(1)_L$ by the background gauge field. This explicit symmetry breaking generates a renormalization group (RG) flow from a conformal field theory in the ultraviolet (UV) to an anisotropic system in the infrared (IR), as the one described by the effective multi-Weyl Hamiltonian in Eq.~\eqref{eq:hmulti}. Therefore, along the RG trajectory the anomalous conductivities are not necessarily anomaly protected. However, in some cases the IR conductivities show a universal behavior~\cite{Amado:2014mla,Copetti:2016ewq}.

In order to understand the one point functions of the currents at very low energies, is necessary to select a specific model. We focus on a strongly coupled theory with holographic dual, considering the simplicity of the computations of one point functions in this case. A second reason for selecting a holographic model is associated to the ambiguities we may encounter related to different regularization schemes~\cite{Landsteiner:2013aba,Grushin:2012mt} in perturbative quantum field theories. Fortunately, in holography these ambiguities are not present due to the existence of a natural regulator (location of the AdS boundary). As a matter of fact, holography has been a vital tool for the understanding of the anomaly-induced transport~\cite{Erdmenger:2008rm,Banerjee:2008th,Rebhan:2009vc,Gynther:2010ed}, and our main goal is to qualitatively demonstrate whether the new predicted transport coefficients survive along the RG flow.

As a first approach to the problem, and for simplicity we only consider the response of the charged current to external gauge fields. We leave the study of chiral vortical conductivites and response in the energy momentum tensor for a future investigation. That allow us to take the probe approximation, in which the bulk gauge fields do not backreact on the geometry. As a consequence the mixed gauge-gravitational anomaly decouples. However, as already seen in previous sections the mixed anomaly is relevant for the chiral vortical or chiral magnetic effects in the energy current.

In holography, the problem of quantum anomalies is well understood and their presence is realized via the introduction of Chern-Simons terms in the bulk action. Our perspective is completely phenomenological, i.e \emph{bottom up}. We assume the existence of a large $N_c$ (color number), strongly coupled gauge theory with a holographic dual and with the same  anomaly structure as in our multi-Weyl system. Therefore, the simplest holographic model we can construct has the form
\begin{eqnarray}
\nonumber S&=&  - \int \mathrm{Tr}\left[\frac{1}{2n}\mathcal  F\wedge {^\star \mathcal F} +  \frac{1}{2c(n)}G\wedge{^\star G}+\right.\\
&&\left. +\lambda\left(A\wedge\mathrm (d A)^2 + \frac{3}{2} A^3\wedge \mathrm d A + \frac{3}{5} A^5\right)\right],\label{eq:HoloAct}
\end{eqnarray}
where the gauge fields are defined as
\begin{equation}
\mathcal A=\mathcal A^{0}s_0\quad,\quad \mathbb{A}=\mathbb{A}^{i}s_i\quad,\quad  A =\mathcal A+\mathbb{A}\,,
\end{equation}
and $s_a=(s_0,s_i)$ are the identity and $SU(2)$ generators introduced in the Sec.~\ref{Sec:Lowenergy_MultiWeyl}. The corresponding field strength associated to the gauge fields are 
\begin{equation}
\mathcal F=d\mathcal A\quad,\quad G=d\mathbb{A} - i \mathbb{A}^2\quad,\quad F = \mathcal F+ G\,.
\end{equation}

We seek to compute the anomalous currents at finite temperature. To do so, we need a finite temperature background geometry which we choose to be the Schwarzschild-AdS blackhole 
\begin{equation}
\mathrm d s^2 = \frac{1}{r^2}\left(-u(r)\mathrm dt^2 + \frac{1}{u(r)}\mathrm dr^2+ dx^2 + dy^2 + dz^2 \right)\,,
\end{equation}
with the horizon at $r_h=1$, and the blackening factor $u(r)=1-r^2$. In these units the Hawking temperature is given by $T=\pi^{-1}$.

To connect the Chern-Simons term in the action Eq. \eqref{eq:HoloAct} with the anomaly, we use the fact that the gauge/gravity duality establishes that the onshell action $S$ corresponds to the boundary QFT effective action $W[A]$, with
\begin{equation}~\label{eq:HolDict}
W[A]=S_{\mathrm{onshell}}[A]\,.
\end{equation}
Therefore, after performing a bulk gauge transformation $\delta_\theta A_M = -\mathcal D_M\theta$, where $M=0,1,\ldots,4$ and $x^M=(t,x,y,z,r)$, we reproduce the following expression
\begin{equation}
\delta_\theta W(A) = \int \mathrm d^4x\, \theta^a G^a[A],
\end{equation}
where $G^a$ is the anomaly [see right-hand side  in Eq. \eqref{eq:consWI}]. This computation shows how the anomaly is realized within the holographic set-up and allows us to fix the coupling $\lambda$,
\begin{equation}
\lambda=\frac{N_c}{24\pi^2} .
\end{equation}
As a next step, we define the holographic one point functions by taking functional derivatives of  Eq. \eqref{eq:HolDict}. In particular, derivatives with respect to the gauge fields generate the (unrenormalized) consistent charged currents
\begin{equation}\label{eq:currentdef}
\tilde J_a^\mu = \frac{\delta S}{\delta  A^a_\mu} = \sqrt{-g}\left(2  F_a^{r\mu} - K^\mu_a[ A]\right)|_\mathrm{boundary}\,,
\end{equation}
where 
\begin{equation}
K^\mu_a[ A] = -\frac{\lambda}{2}\epsilon^{\mu\nu\rho\sigma}\mathrm{Tr}\left[ s_a\left( \{A_\nu,  F_{\rho\sigma}\} -  A_\nu A_\rho A_\sigma \right)\right]
\end{equation}
is precisely  the Bardeen-Zumino polynomial. Consequently, we can read from Eq. \eqref{eq:currentdef}  the holographic definition of the (unrenormalized) covariant current
\begin{equation}\label{eq:covcurrentdef}
J_a^\mu = 2\sqrt{-g} \: F_a^{r\mu}|_\mathrm{boundary}.
\end{equation}

After defining the model, we introduce the bulk gauge field ansatz, which is dual to the boundary field theory at finite density and temperature. Furthermore, in order to study the anomaly-induced currents, we introduce an external background magnetic field ${\bf B}=(0,0,B)$ and the symmetry breaking non-Abelian gauge field responsible for the multi-Weyl spectrum in the weakly coupled model. Thus, the bulk gauge field at the boundary has to take the value
\begin{equation}\label{eq:boundcond}
 A(r_b) = \left(\mathcal \mu s_0 +\mathcal \mu_3s_z\right)\mathrm d t +  \Delta\left(s_x\mathrm dx +s_y\mathrm dy\right)  + x B s_0\mathrm d y\,.
\end{equation}
With all these ingredients, the simplest ansatz we need to consider takes the following form
\begin{eqnarray}~\label{eq:ansatz}
\nonumber  A(r) &=& \left( A_t(r)s_0 + A_t^3(r)s_z\right)\mathrm d t + \mathcal Q(r)\left(s_x\mathrm dx +s_y\mathrm dy\right) + \\
&& + \left( A_z(r)s_0 + A_z^3(r)s_z\right)\mathrm d z + x Bs_0\mathrm d y\,.
\end{eqnarray}
This ansatz needs to be plugged into the equations of motion and solved imposing the boundary conditions Eq. \eqref{eq:boundcond} at the boundary and regularity in the interior of the  space-time except for the fields $A_t$ and $A^3_t$, which has to vanish. In the appendix~\ref{append:EoM} we show the explicit form of the equations of motion.

\begin{figure}[t!]
	\centering
	\includegraphics[width=0.4\textwidth]{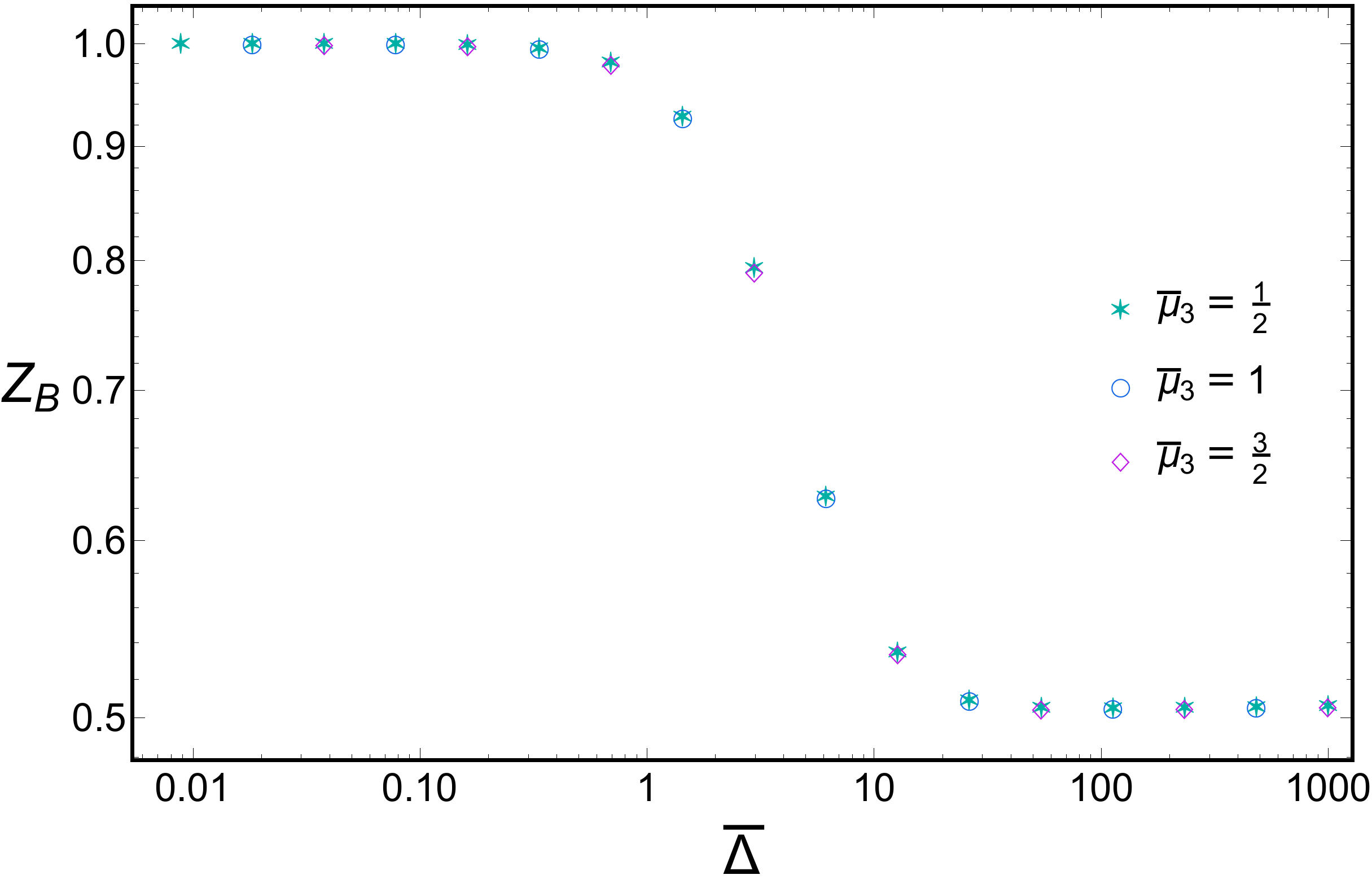}
	\caption{Isospin chiral magnetic conductivity as a function of $\bar \Delta$ for vanishing chemical potential ($\mu$).}
	\label{fig:ZB}
\end{figure}


A first conclusion we make after writing down the equations of motion, is that the Abelian covariant current can be obtained analytically and takes its universal form
\begin{eqnarray}
J^z  &=& \frac{n}{4\pi^2}\mu B+ \frac{c(n)}{4\pi^2}\mu_3 \Delta^2 \,.
\end{eqnarray}
On the other hand, the non-Abelian sector needs to be solved numerically.

Before showing the results for the non-Abelian current we discuss some aspects of the boundary sources $\mu,\mu_3$ and $\Delta$. They introduce a deformation of the UV conformal field theory as follows
\begin{equation}
\mathcal L=\mathcal L_{\rm CFT} + \mu Q+ \mu_3 Q^3 + \Delta\left(\delta^\mu_xj^1_\mu+\delta^\mu_yj^2_\mu\right)\,.
\end{equation}
In particular, the charge $Q$ remains conserved (modulo the anomaly). However, by the presence of $\Delta$, all the charges $Q^i$ are not conserved, because the $SU(2)$ group is explicitly broken. Hence, $\Delta$ has to be understood as a coupling constant, which gets renormalized along the RG flow, since no symmetry protects its value. For that reason, we define the IR renormalized $\Delta$ as follows
\begin{equation}
\Delta_{\rm IR}=\mathcal Q(r_h).
\end{equation}
At this point it is convenient to introduce the set of adimensional variables
\begin{equation}
\bar \Delta = \frac{\Delta}{T} \quad,\quad\bar \mu = \frac{\mu}{T}
\quad,\quad\bar \mu_3 = \frac{\mu_3}{T}.
\end{equation}

In Fig.~\ref{fig:bIR}, we show the renormalized $\Delta_{\rm IR}$ for several values of the chemical potentials, and find it to be independent of $\bar{\mu}$ and $\bar{\mu}_3$. Therefore, we infer that $\Delta_{\rm IR}=T Z(\bar \Delta)$ with 
\begin{equation}~\label{Eq:ZDelta}
Z(\bar \Delta) \approx \left\{\begin{array}{lcr}
\bar \Delta &\mathrm{for} & \bar \Delta\ll 1 \\
5.7584 & \mathrm{for} &\bar \Delta \gg 1
\end{array}\right.\,.
\end{equation} 
From the above dependencies, we conclude that $\Delta_{\rm IR}\sim 5.8 T$ for small enough temperatures.

\begin{figure}[t!]
	\centering
	\includegraphics[width=0.4\textwidth]{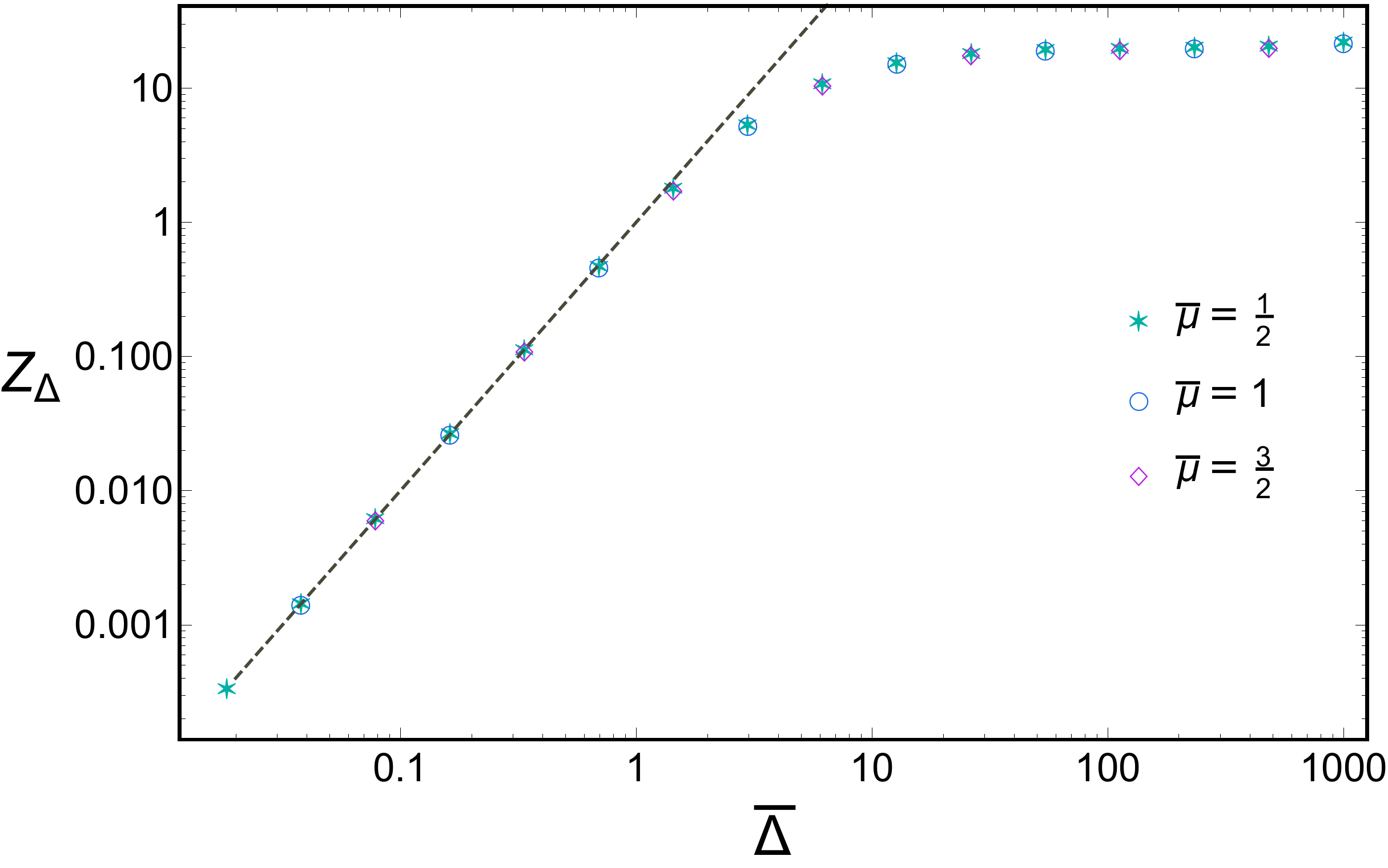}
	\caption{Isospin current as a function of $\bar \Delta$ for vanishing isospin chemical potential ($\mu_3$). The dashed line corresponds with a a quadratic fitting $Z_\Delta= \bar \Delta^2$.}
	\label{fig:Zb}
\end{figure}

We now compute the isospin current ($J_3^z$) either in the presence of isospin chemical potential ($\mu_3$) or chemical potential ($\mu$). In the former case, we found the following generic behavior
\begin{equation}
J_3^z =Z_B\left(\bar \Delta\right) \frac{c(n)\mu_3}{4\pi^2} B \,,
\end{equation}
and the function $Z_B$ appears to be independent of $\mu_3$, as can be seen in Fig.~\ref{fig:ZB}. Numerical analysis suggest that $Z_B(\bar \Delta)$ has the following functional form
\begin{equation}
Z_B(\bar \Delta) \approx \left\{\begin{array}{lcr}
1 & \mathrm{for}& \bar \Delta\ll 1 \\
0.506 & \mathrm{for} &\bar \Delta \gg 1
\end{array}\right.\,,
\end{equation} 
which implies that at low temperatures the isospin chiral magnetic effect is reduced by a factor of $2$ (approximately), yielding
\begin{equation}~\label{j3z_1}
J_3^z \approx 0.5 \frac{n}{4\pi^2} \mu_3B \,.
\end{equation}
Next we study the system only in presence of the Abelian chemical potential $\mu$. In this case, the isospin current takes the following form
\begin{equation}
J_3^z =   Z_{\Delta}\left(\bar \Delta\right)\frac{c(n)}{4\pi^2}\mu T^2\,.
\end{equation}
The functional dependence of $Z_\Delta$ on $\bar{\Delta}$ is shown in Fig. \ref{fig:Zb}, and it depends only on the symmetry breaking parameter $\bar \Delta$. In the two asymptotic regimes, it is characterized as follows
\begin{equation}
Z_\Delta(\bar \Delta) \approx \left\{\begin{array}{lcr}
\bar \Delta^2 & \mathrm{for} &\bar \Delta\ll 1 \\
21.762 & \mathrm{for} &\bar \Delta \gg 1
\end{array}\right.\,.
\end{equation} 
As expected at high enough temperatures the current takes the universal value fixed by the anomaly 
\begin{equation}
J_3^z \approx   \frac{c(n)}{4\pi^2}\mu \Delta^2\,.
\end{equation}
However, at very low temperatures it has a different behavior, given by
\begin{equation}
J_3^z \approx   21.8\frac{c(n)}{4\pi^2}\mu T^2\,,
\end{equation}
which is set by the temperature instead of the parameter $\Delta$. Since, the renormalized IR value of the non-Abelian gauge field is $\Delta_{\rm IR}\approx 5.8T$ [see Eq.~(\ref{Eq:ZDelta})], the current can alternatively be written as
\begin{equation}~\label{j3z_2}
J_3^z \approx    0.6 \; \frac{c(n)}{4\pi^2} \; \mu \; \Delta_{\rm IR}^2\,.
\end{equation}
At this point a comment is due regarding the probe approximation. Ignoring the backreaction of the gauge fields on the space-time geometry is a valid approximation, as long as the temperature is not `too low'. Therefore, some numerical deviations of the coefficients [see Eqs.~(\ref{j3z_1}) and (\ref{j3z_2})] can be expected for the IR conductivities.

 Therefore, at this point we cannot conclude regarding the universalities associated with the conductivities. Notheless, the important fact is that their coefficients are of order one, which strongly suggest that even in the IR low energy model the anomalous transport survive in the isospin current, which is directly connected to the $SO(2)$ rotational invariance of the low-energy sector.

\section{Discussions and future directions}~\label{Sec:Discussion}

The primary goal of the present work is to develop a comprehensive understanding of anomalies in multi-Weyl semimetals. The lack of the Lorentz symmetry (stemming from the nonlinear band dispersion) yields complex and rich structure in the theory, which motivates us to find a suitable generalization of these models in order to shed light on the desired physical phenomena. The appropriate formulation is presented in Sec.~\ref{Sec:Lowenergy_MultiWeyl}, unveiling the presence of an underlying non-Abelian symmetry and associated anomalies, which were hitherto missed. This construction is based on $n$ copies of simple Weyl fermions, coupled via a particular spin-orbit coupling that preserves certain discrete crystalline symmetry, but breaks the Lorentz invariance and couples with linearly dispersing chiral fermions as a static $SU(2)$ gauge field. As a result, the multi-Weyl spectra are recovered at low energies. Our study aims at a systematic analysis of mixed Abelian and non-Abelian anomalies and associated transport, which we achieve by employing various approaches available in condensed matter and high-energy physics.

We introduce both two band models for general Weyl semimetals (with $n=1,2,3$), as well as the four (six) band model for the double (triple) Weyl semimetals on a cubic lattice in Sec.~\ref{subsec:onlylatticemodels}. These models produce the correct low-energy descriptions for multi-Weyl semimetals (compare Figs.~\ref{Fig:Bandstructure_2band} and \ref{Fig:Bandstructure_multiband}), and the bulk-boundary correspondence, encoded in the number ($n$) of topologically protected Fermi arc surface states (see Sec.~\ref{subsec:fermiarc} and Figs.~\ref{Fig:Fermiarcs_2band} and \ref{Fig:Fermiarcs_multiband}). At this point we make an interlude in the lattice simulations to construct an effective field theory for multi-Weyl systems.

The key observation while arriving at the effective field theory is the presence of a non-Abelian $U(2)$ flavor symmetry, which is
controlled by a background field and responsible for the multi-Weyl dispersion at low energies. This representation allows us to write a Lagrangian in the form of free fermions coupled to a non-Abelian background field. Subsequently, we employ powerful techniques of quantum field theory to identify Abelian and non-Abelian currents that exhibit anomalous nonconservation due to pertinent gauge and gravitational anomalies. Such an elegant formulation allows us to relate the anomaly structures in multi-Weyl semimetals with the ones previously known for Lorentz symmetric systems, and at the same time to bypass cumbersome computational steps. Most importantly, we manage to extract possible transport contributions that stem from anomalies (such as non-Abelian generalization of anomalous Hall conductivity) and are \emph{unique} to multi-Weyl systems.

We confirm some of the field theoretic predictions from the lattice models and the gauge/gravity duality. In particular, we numerically compute the regular and isospin charge accumulations respectively as a function of the Abelian and non-Abelian magnetic fields for both double and triple Weyl semimetals. Remarkably, the numerical outcomes display excellent agreement with the field theory, at least when the field strengths are sufficiently weak, see Figs.~\ref{Fig:AbelianAnomaly_Summary} and \ref{Fig:nonAbelian_summary}. Finally we employ the holographic techniques to study a strongly coupled version of multi-Weyl semimetals. From holography, we analyze the corresponding renormalization group flow in a particular theory that is dual to the Maxwell-Einstein system with a non-Abelian Chern-Simons term, and show that the non-Abelian anomalous transport coefficients even though get renormalized, remain finite in the infrared regime.

There are many interesting future outgrowths of the current investigation. We now highlight some of the most exciting ones. Our theoretical analysis suggests possible experimental ramifications of non-Abelian anomaly in magnetotransport. In order to capture its signatures on magnetoresistance, one needs to systematically develop a semiclassical framework by taking into account the non-Abelian currents, see for example Refs.~\cite{PhysRevLett.51.351,LITIM2002451}.

Yet another avenue to explore is to demonstrate/verify field theoretic predictions from concrete lattice models. In this work, we introduce only the simple setup with a background non-Abelian field, and compute the regular and isospin charge accumulation in the system. For example, in the same setup one can also introduce chiral chemical potential by adding a term $\mu_5 \sin(k_z a)$ in the lattice model. Furthermore, one can perform detailed numerical investigations in the presence of axial magnetic fields (obtained via local deformations of hopping parameters) or gravitational fields in multi-Weyl semimetals.

We also note that the transport coefficients associated to the non-Abelian current get renormalized, and thus are generically different in the infrared and ultraviolet regimes. Therefore, a detailed field theoretic analysis leading to the renormalization group flow of these coefficients using the Feynman diagrammatic expansion is due, which we leave for a future investigation.  

Our field theoretic analysis can be extended for multi-fold fermions~\cite{2016arXiv160303093B,2017Natur.547..298B}, for which the irreducible band representation transform under spin-$S$ representation, where $S$ can be half-integer or an integer. At the Hamiltonian level this can be accomplished by replacing two-dimensional Pauli matrices (${\boldsymbol \tau}$) by $(2 S+1)$-dimensional spin-$S$ matrices. A systematic derivation of the effective field theory for multi-fold fermions is left for a future investigation. Finally, we can extend the holographic studies of a multi-Weyl systems to compute the vortical conductivities and energy-current, for which the backreaction from the gauge field on the metric should accounted for.

\acknowledgments

F. P-B. and P. S. acknowledge useful conversations with Karl Landsteiner and Maria Vozmediano, and hospitality of Nordita. P. S. was supported by the Deutsche Forschungsgemeinschaft via the Leibniz Programm. F.P.-B., P.S. and R.M.A.D. Aknowledge DFG through Würzburg-Dresden Cluster of Excellence on Complexity and Topology in Quantum Matter - ct.qmat (EXC 2147, project-id 39085490). F.P.-B. acknowledges the Quantum Matter Academy of ct.qmat for support.

\appendix

\section{Low-energy Hamiltonian for multi-Weyl fermions}~\label{appendix:lowenergymodel}

We devote this appendix to derive the low-energy Hamiltonian for double and triple Weyl semimetals [see Eq.~(\ref{eq:hmulti})] starting from coupled two and three copies of simple Weyl fermions [see Eq.~(\ref{Eq:genWeyl_coupled})], respectively. Accordingly, we introduce a four and six component spinors
\begin{eqnarray}
\Psi^\top({\bf p})_{\rm DW} &=& \left[ c_{1, \uparrow}, c_{1, \downarrow}, c_{2, \uparrow}, c_{2, \downarrow}\right] ({\bf p}), \nonumber \\
\Psi^\top({\bf p})_{\rm TW} &=&\left[ c_{1, \uparrow}, c_{1, \downarrow}, c_{2, \uparrow}, c_{2, \downarrow}, c_{3, \uparrow}, c_{3, \downarrow} \right] ({\bf p}),  
\end{eqnarray}
for these two systems, where $c_{j, \tau}({\bf p})$ is the fermionic annihilation operator with pseudospin projection $\tau=\uparrow, \downarrow$, flavor index $j=1,2,3$ and momenta ${\bf p}$, measured from the Weyl node, located at ${\bf p}=0$. In this basis, the Hamiltonian operators yielding double and triple Weyl fermions at low energies respectively take the form
\begin{widetext} 
\begin{eqnarray}
H_{\rm DW} &=& \left[ \begin{array}{c|c}
v_\perp \left({\boldsymbol \tau}_\perp \cdot {\bf p}_\perp \right) + v \tau_z p_z & \Delta \left( \tau_1 -i \tau_2 \right)/2 \\
\hline
\Delta \left( \tau_1 +i \tau_2 \right)/2 & v_\perp \left({\boldsymbol \tau}_\perp \cdot {\bf p}_\perp \right) + v \tau_z p_z
\end{array}
\right], \\
H_{\rm TW} &=& \left[ \begin{array}{c|c|c}
v_\perp \left({\boldsymbol \tau}_\perp \cdot {\bf p}_\perp \right) + v \tau_z p_z & \Delta \left( \tau_1 -i \tau_2 \right)/2 & \hat{0}_{2\times 2} \\
\hline
\Delta \left( \tau_1 +i \tau_2 \right)/2 & v_\perp \left({\boldsymbol \tau}_\perp \cdot {\bf p}_\perp \right) + v \tau_z p_z & \Delta \left( \tau_1 -i \tau_2 \right)/2 \\
\hline
\hat{0}_{2\times 2} & \Delta \left( \tau_1 +i \tau_2 \right)/2 & v_\perp \left({\boldsymbol \tau}_\perp \cdot {\bf p}_\perp \right) + v \tau_z p_z
\end{array}
\right],
\end{eqnarray}  
\end{widetext}
where ${\bf p}_\perp= (p_x,p_y)$, ${\boldsymbol \tau}_\perp=(\tau_1, \tau_2)$ and $\hat{0}_{2 \times 2}$ represents a two-dimensional null matrix. Note that for ${\bf p}=0$, $c_{1,\downarrow}$ and $c_{2,\uparrow}$ degrees of freedom are gapped and placed at energies $\pm \Delta$ for double-Weyl system. For triple-Weyl system $c_{1,\downarrow}$ and $c_{3,\uparrow}$ ($c_{2,\uparrow}$ and $c_{3,\downarrow}$) are placed at energy $+(-)\Delta$. Hence, these degrees of freedom do not participate (approximately) in the low-energy dynamics of the multi-Weyl systems. We, therefore, integrate them out in order to arrive at the effective low-energy models. If we denote the spinor basis for low- and high-energy degrees of freedom for double [triple] Weyl system as $\Psi_{\rm DW, L} ({\bf p}) [\Psi_{\rm TW,L} ({\bf p})]$ and $\Psi_{\rm DW,H} ({\bf p}) [\Psi_{\rm TW,H} ({\bf p})]$, respectively then 
\begin{eqnarray}
\Psi^\top_{\rm DW, L} ({\bf p}) &=& \left[ c_{1,\uparrow}, c_{2,\downarrow}\right]({\bf p}), \:
\Psi^\top_{\rm TW, L} ({\bf p})  =  \left[ c_{1,\uparrow}, c_{3,\downarrow}\right]({\bf p}), \nonumber \\
\Psi^\top_{\rm DW, H} ({\bf p}) &=& \left[ c_{1,\downarrow}, c_{2,\uparrow}\right]({\bf p}), \\
\Psi^\top_{\rm TW, H} ({\bf p}) &=& \left[ c_{1, \downarrow}, c_{2, \uparrow}, c_{2, \downarrow}, c_{3, \uparrow} \right]({\bf p}). \nonumber 
\end{eqnarray} 
One can integrate out the split off bands in the path integral formalism in the following way. 

The imaginary time partition function for multi-Weyl system reads 
\begin{widetext}
\begin{eqnarray}
Z_a &=& \int {\mathcal D} \Psi^\dagger_{a,L} {\mathcal D} \Psi_{a,L} \: \exp\left[ -\int d\tau \; \Psi^\dagger_{a,L} \left( \partial_\tau + H^{a}_{LL} \right) \Psi_{a,L} \right] 
\int {\mathcal D} \Psi^\dagger_{a,H} {\mathcal D} \Psi_{a,H} \nonumber \\
&\times& \exp\left[ -\int d\tau \left\{ \Psi^\dagger_{a,H} \left( \partial_\tau + H^{a}_{HH} \right) \Psi_{a,H}  
+ \Psi^\dagger_{a,L} H^{a}_{LH} \Psi_{a,H} + \Psi^\dagger_{a,H} H^{a}_{HL} \Psi_{a,L} \right\} \right],
\end{eqnarray}
where $a={\rm DW}, {\rm TW}$, and 
\begin{eqnarray}
H^{\rm DW}_{LL} &=& \left[ \begin{array}{cc}
v p_z & 0 \\
0 & -v p_z
\end{array}\right]= H^{\rm TW}_{LL}, \: 
H^{\rm DW}_{HH}= \left[ \begin{array}{cc}
-v p_z & \Delta \\
\Delta & v p_z
\end{array}\right], \:
H^{\rm DW}_{LH}= \left[ \begin{array}{cc}
f({\bf p}_\perp) & 0 \\
0 & f^\ast({\bf p}_\perp)
\end{array}\right], \:
H^{\rm DW}_{HL}=\left( H^{\rm DW}_{LH} \right)^\dagger, \nonumber \\
H^{\rm TW}_{HH} &=& \left[ \begin{array}{cccc}
- v p_z & \Delta & 0 & 0 \\
\Delta & v p_z & f({\bf p}_\perp) & 0 \\
0 & f^\ast({\bf p}_\perp) & - v p_z & 0 \\
0 & 0 & \Delta & v p_z
\end{array}
\right], \:
H^{\rm TW}_{LH}= \left[ \begin{array}{cccc}
f({\bf p}_\perp) & 0 & 0 & 0 \\
0 & 0 & 0 & f^\ast({\bf p}_\perp)
\end{array}
\right]. 
\end{eqnarray}
In the above expressions $f({\bf p}_\perp)= v \left( p_x - i p_y \right)$. Upon integrating out the high-energy degrees of freedom we arrive at the renormalized partition function for the low-energy modes 
\begin{eqnarray}
Z^L_{a}= \int {\mathcal D} \Psi^\dagger_{a,L} {\mathcal D} \Psi_{a,L} \: \exp\left[ -\int d\tau \Psi^\dagger_{a,L} \left( \partial_\tau + H^{a}_{LL} - H^{a} G^a_{HH}(i \omega_n) H^a_{HL} \right) \Psi_{a,L} \right],
\end{eqnarray}
where $G^a_{HH}(i \omega_n)=\left( i \omega_n - H^a_{HH} \right)^{-1}$ and $\omega_n$ is the Mastubara frequency. Setting $\omega_n=0$, from the above expression we arrive at the renormalized Hamiltonian ($H^{a, {\rm Ren}}_{LL}$) in terms of the low-energy modes
\begin{equation}
H^{a, {\rm Ren}}_{LL}= H^{a}_{LL} - H^{a} G^a_{HH}(0) H^a_{HL}.
\end{equation}    
After some lengthy, but straightforward algebra we find 
\begin{eqnarray}
H^{{\rm DW}, {\rm Ren}}_{LL} &=& v p_z \tau_3 - \frac{v^2_\perp \Delta}{\Delta^2+ v^2 p^2_z} \; \left[ (p^2_x-p^2_y) \tau_1 + 2 p_x p_y \tau_2  \right] + v p_z \left[ \frac{v^2_\perp k^2_\perp}{\Delta^2 + v^2 p^2_z} \right] \tau_0, \\
H^{{\rm TW}, {\rm Ren}}_{LL} &=& v p_z \tau_3 + \frac{v^3_\perp \Delta^3}{\left[ (\Delta^2 + v^2 p^2_z)^2 - v^2 v^2_\perp k^2_z k^2_\perp \right]} \; \left[ p_x(p^2_x-3p^2_y) \tau_1 + p_y(p^2_y-3p^2_x) \tau_2 \right] \nonumber \\
&+& v p_z \; \left[ v^2_\perp p^2_\perp \: \frac{v^2_\perp p^2_\perp- v^2_z p^2_z-\Delta^2}{(\Delta^2 + v^2 p^2_z)^2 - v^2 v^2_\perp k^2_z k^2_\perp}\right] \tau_0.
\end{eqnarray}
\end{widetext}
Note that the particle-hole asymmetric terms (proportional to $\tau_0$) vanish at the Weyl nodes, located at ${\bf p}=0$. Then for $\Delta \gg v_\perp |{\bf p}_\perp|, v p_z$ we arrive at the low-energy Hamiltonian for double and triple Weyl semimetal with $\alpha_n=v^n_\perp/\Delta^{n-1}$ respectively for $n=2$ and $3$, see Sec.~\ref{Sec:Lowenergy_MultiWeyl}.

\section{Chiral vortical conductivities}~\label{Append:vorticalnonAb}

In this appendix we display the transport coefficients omitted in Sec.~\ref{Sec:QFT_MultiWeyl}. In particular, the chiral vortical effect in the Abelian currents is given by
\begin{eqnarray}
\mathbf J_e &=& \left[ \frac{n}{2\pi^2} \; \mu  \; \mu_5 + c(n) \; \mu_3 \; \mu_{3_5}  \right] {\boldsymbol \omega}\,, \\
\nonumber\mathbf J_5 &=& \left[ \frac{n}{4\pi^2}\left(\mu ^2+\mu_5^2+\frac{\pi^2T ^2}{3} \right) + \frac{c(n)}{4\pi^2} \; \left(\mu_3 ^2+\mu_{3_5}^2 \right) \right] {\boldsymbol \omega} \,,\\
\end{eqnarray}
whereas in the non-Abelian currents it takes the form
\begin{eqnarray}
\mathbf J_3 &=& \frac{c(n)}{2\pi^2} \: \big( \mu \; \mu_{3_5} + \mu_3 \; \mu_5 \big) \; {\boldsymbol \omega}\,, \\
\mathbf J_{3_5} &=&\frac{c(n)}{2\pi^2} \: \big( \mu \; \mu_{3} + \mu_{3_5} \; \mu_5 \big) \; {\boldsymbol \omega}\,.
\end{eqnarray}
On the other hand, the energy current generated by the chiral magnetic and vortical effects read
\begin{widetext}
\begin{eqnarray}
{\mathbf J}_\epsilon &=& \frac{1}{2\pi^2} \; \left[ c(n) \; \mu_3 \; \mu_{3_5} + n \; \mu \; \mu_5 \right] {\mathbf B} + \frac{c(n)}{2\pi^2} \; \left[ \mu \; \mu_{3_5} + \mu_3 \; \mu_5 \right] \; {\mathbf B}_3 + 
\nonumber \frac{1}{4\pi^2} \; \left[ c(n) \; \left( \mu_3^2+\mu_{3_5}^2\right) + n \left(\mu ^2+\mu_5^2\right)+\frac{n \pi^2T ^2}{3}\right] \; {\mathbf B}_5 \nonumber \\
&+&  \frac{c(n)}{2\pi^2} \left[ \mu \; \mu_3 + \mu_5 \; \mu_{3_5} \right] \; \mathbf B_{3_5} + \frac{n}{6\pi^2} \; \left[ \mu_5 \; (3\mu^2+\mu_5^2) + \mu_5 \; \pi^2 T^2 \right] \; {\boldsymbol \omega}
+ \frac{c(n)}{2\pi^2} \; \left[ 2 \mu \; \mu_3 \; \mu_{3_5} + \mu_5 \; (\mu_3^2+\mu_{3_5}^2) \right] {\boldsymbol \omega}.
\end{eqnarray}
\end{widetext}


\section{Equations of motion}~\label{append:EoM}

In this appendix, we show the equations of motion for the holographic theory, introduced in Eq.~(\ref{eq:HoloAct}). The Maxwell-Yang-Mills-Chern-Simons equations on the curved background are
\begin{eqnarray}
\mathcal \nabla_\mu \mathcal F^{\mu\nu} - 6\lambda\epsilon^{\nu\rho\alpha\beta\gamma}\mathrm{Tr}\left( F_{\rho\alpha}F_{\beta\gamma} \right) &=& 0, \\
\mathcal D_\mu G^{a,\mu\nu} - \frac{3}{2}\lambda\epsilon^{\nu\rho\alpha\beta\gamma}\mathrm{Tr}\left( s^aF_{\rho\alpha}F_{\beta\gamma} \right) &=& 0\,.
\end{eqnarray}
After evaluating the ansatz Eq.~(\ref{eq:ansatz}), we find a set of equations for the Abelian and non-Abelian fields. First we show the equations of motion for the Abelian sector
\begin{eqnarray}
\left(\frac{A_t '}{r} + 24\lambda\left[c(n) \mathcal Q^2 A_z^3 +nB A_z\right]\right)' &=& 0, \\
\left(\frac{u(r) A_z'}{r} + 24 \lambda\left[  c(n)\mathcal Q(r)^2 A^3_t+ n B A_t\right] \right)'  &=& 0\,,
\end{eqnarray}
where the prime symbol stands for $\partial_r$. Notice that this equations can be integrated once, leading to 
\begin{eqnarray}
\frac{A_t '}{r} &=& 4\rho - 24\lambda\left[c(n) \mathcal Q^2 A_z^3 +nB A_z\right]\,, \\
\frac{u(r) A_z'}{r} &=& -24 \lambda\left[  c(n)\mathcal Q(r)^2 A^3_t+ n B A_t\right] \,,
\end{eqnarray}
where $\rho$ is an integration constant, and the regularity at the horizon for $A_z$ forbids the presence of an extra integration constant.
 Now we show the equations for the non-Abelian sector, where the symmetry breaking field $\mathcal Q(r)$ satisfies
\begin{equation}
\left(\frac{u(r) \mathcal Q'}{r} \right)' +  \left(\frac{(A_t^3)^2}{r u(r)}-\frac{(A_z^3)^2}{r}-\frac{\mathcal  Q^2}{r}\right)\mathcal Q =0\,.
\end{equation}
The rest of the equations read 
\begin{eqnarray}
\nonumber\left(\frac{u(r) A_z^{3'}}{r} \right)'  -\frac{2 \mathcal Q^2 A_z^3}{r } +24 \lambda  c(n) \left( \mathcal Q^2 A_t'(r)+ B A_t^{3'}\right) &=& 0 \,,\\
&&\\
\nonumber\left(\frac{A_t^{3 '}}{r} \right)' -\frac{2 \mathcal Q^2 A_t^3}{r u(r)}+24 \lambda  c(n) \left( \mathcal Q^2 A_z'(r)+ B A_z^{3'}\right)&=& 0\,.\\
\end{eqnarray}

\bibliography{Anomaly_Draft_V1_2}

\end{document}